\documentclass[onecolumn,prd，eprint,aps,groupedaddress,amsmath,amssymb,11pt,a4paper]{revtex4-2}

\usepackage{graphicx}
\graphicspath{{v2/}}
\usepackage{subfigure} 
\usepackage{bm}
\usepackage{amssymb}
\usepackage{amsmath}
\usepackage{epsfig}
\usepackage{cancel}
\usepackage{color}
\usepackage{mathtools}
\usepackage{cases}
\usepackage{booktabs}
\usepackage{physics}
\usepackage{braket}
\usepackage{placeins}

\usepackage{tikz}
\usepackage{tikz-3dplot}

\usepackage{float}     
\usepackage{multirow}
\usepackage{bbm}
\usepackage{comment}
\usepackage{natbib}
\usepackage{array}
\usepackage{dcolumn}

\definecolor{blue}{rgb}{0,0,0} 
\definecolor{linkblue}{rgb}{0,0.3,0.8}  
\definecolor{citecyan}{rgb}{0,0.7,0.9}  

\usepackage[
colorlinks=true,
filecolor=black,
anchorcolor=linkblue,
linkcolor=linkblue,
citecolor=citecyan,
urlcolor=linkblue,
linktocpage=true,
plainpages=false,
breaklinks=true,
pdfstartview=FitH
]{hyperref}

\renewcommand*{\d}{\ \mathrm{d}}
\newcommand*{\diff}[2]{\frac{\d #1}{\d #2}}

\renewcommand*{\i}{\mathrm{i}}
\newcommand*{\e}{\ \mathrm{e}}

\newcolumntype{C}[1]{>{\centering\arraybackslash}p{#1}}

\begin{document}

\title{Accretion Disk Perturbations and Their Effects on Kerr Black Hole Superradiance and Gravitational Atom Evolution}

\author{Ruiheng Li}
\email{lirh58@mail2.sysu.edu.cn}
\author{Zhong-hao Luo}
\email{luozhh58@mail2.sysu.edu.cn}
\author{Zehong Wang}
\author{Fa Peng Huang}
\email{huangfp8@sysu.edu.cn}
\thanks{Corresponding author.}

\affiliation{MOE Key Laboratory of TianQin Mission,
TianQin Research Center for Gravitational Physics \& School of Physics and Astronomy,
Frontiers Science Center for TianQin,
Gravitational Wave Research Center of CNSA,
Sun Yat-sen University (Zhuhai Campus), Zhuhai 519082, China}

\
\begin{abstract}
Kerr black hole (BH) superradiance can form gravitational atoms and produce characteristic gravitational-wave signals, providing a probe of ultralight bosons and dark matter. In realistic systems, accretion-disk gravity can shift energy levels and mix states, modifying the effective superradiant growth. We model the disk as a weak external perturbation via a multipole expansion and derive an effective three-level Hamiltonian for the $n=2$ subspace $\{\ket{211},\ket{210},\ket{21-1}\}$ in the weak-coupling regime. The leading disk effect is the quadrupolar ($\ell_d=2$) tidal field, whose symmetries fix the selection rules: axisymmetry gives only diagonal shifts, equatorial nonaxisymmetry activates $\Delta m=\pm2$ mixing ($\ket{211}\leftrightarrow\ket{21-1}$), and breaking equatorial reflection opens $\Delta m=\pm1$ couplings involving $\ket{210}$. As illustrations, a transient equatorial $m=2$ spiral wave drives the resulting two-level system and can suppress  superradiance by populating a decaying mode, while a quasi-static warp produces full three-level mixing and can generate narrow ``growth gaps'' near accidental near-degeneracies, with the same static reshuffling also allowing enhancement when weight shifts toward the growing mode. These findings demonstrate that accretion disk perturbations are a crucial environmental factor in determining the dynamics of BH superradiance and the evolution of boson clouds, thereby providing a more reliable theoretical basis for assessing the detectability of ultralight bosons in realistic astrophysical settings.
\end{abstract}

\date{\today}
\maketitle

\newpage
\tableofcontents
\allowdisplaybreaks
\newpage

\section{Introduction}
\label{sec:introduction}

In recent years, ultralight bosons have attracted widespread attention as promising dark matter candidates, and Kerr black holes (BHs) provide a unique pathway to test such new physics. If a light scalar field exists around a rotating BH, then under the superradiance condition, incident waves can extract spin energy from the BH and become amplified~\cite{Starobinskii1973AmplificationOW,Brito:2015oca,Press:1972zz,Bardeen:1972fi,Misner:1972kx,Penrose:1971uk,Zouros:1979iw,1971JETPL..14..180Z,Khlopov:1985fch}. A fraction of the amplified waves may be trapped in the gravitational potential well, forming a macroscopically observable boson cloud. Together with the Kerr BH, this constitutes the so-called ``gravitational atom''~\cite{Baumann:2018vus,Arvanitaki:2010sy,Detweiler:1980uk,Cardoso:2004nk,Dolan:2007mj,Baumann:2019eav,PhysRevD.83.044026,Brito:2015oca}. This mechanism not only alters the spin evolution of BHs but may also generate characteristic gravitational wave signals, offering a core avenue to probe new particles in strong-gravity environments~\cite{Brito:2015oca,Arvanitaki:2014wva,Brito2020,Yoshino:2013ofa,Brito:2017zvb}.

\color{blue}
Most early studies of superradiant instabilities~\cite{Arvanitaki:2010sy,Arvanitaki:2014wva,Brito2020,Yoshino:2013ofa,Brito:2017zvb,Baumann:2018vus,Detweiler:1980uk,Cardoso:2004nk,Dolan:2007mj,Baumann:2019eav,PhysRevD.83.044026} focused on isolated BH backgrounds. In recent years, environmental effects~\cite{Tong:2022bbl,Baumann:2019ztm,Xie:2022uvp,Takahashi:2021yhy,Takahashi:2024fyq,Baumann:2021fkf,Baumann:2022pkl,1981ARA&A..19..137P,Page:1974he,Bardeen:1975zz,Brito:2014wla,Brito:2015oca,Sarmah:2024nst,Du:2022trq} on BH superradiance have also been investigated, and these effects arise through several physically distinct mechanisms. Besides the secular spin evolution caused by accretion~\cite{Brito:2014wla,Brito:2015oca,Sarmah:2024nst}, the gravitational field of surrounding matter can directly perturb the superradiant cloud. Static or slowly varying structures, such as accretion disks and stellar halos, mainly modify the cloud spectrum and mix superradiant and decaying eigenlevels~\cite{Du:2022trq}. By contrast, orbiting perturbers---including binary companions and stars or compact objects on inspiralling trajectories such as extreme-mass-ratio inspirals (EMRIs)---generate explicitly time-dependent tidal fields~\cite{Tong:2022bbl,Baumann:2019ztm,Xie:2022uvp,Takahashi:2021yhy,Takahashi:2024fyq,Baumann:2021fkf,Baumann:2022pkl,1981ARA&A..19..137P,Page:1974he,Bardeen:1975zz,Du:2022trq}. Their orbital harmonics can resonantly drive transitions between cloud levels when an integer multiple $k\Omega$ of the orbital frequency matches the corresponding level splitting; for EMRIs, the orbital frequency $\Omega$ itself evolves during the inspiral, so the resonance is swept through rather than maintained. Our work focuses on disk-origin perturbations with specific non-axisymmetric multipoles. The novelty is that realistic disk structures, such as an $m=2$ spiral or an $m=1$ warp, provide definite angular selection channels and can mix growing and decaying gravitational-atom levels when the disk frequency or warp precession matches the intrinsic level splitting.
\color{black}

Motivated by this, we consider thin accretion disks, and study how thin-disk~\cite{Shakura:1973boa,Page:1974he,novikov1973astrophysics} gravity perturbs the gravitational atom during the \emph{linear} superradiant growth stage. We first justify a separation of timescales in which the superradiant growth of the dominant level is faster than the accretion-driven secular drift of the Kerr parameters, allowing $M$ and $\tilde a$ to be treated as effectively constant. We then model the disk through its gravitational potential and develop a perturbative multipole expansion in the cloud region. {\color{blue} In a freely falling frame centered on the BH--cloud system, the monopole and dipole terms do not induce state mixing, so the quadrupole term, $\ell_d=2$, gives the leading contribution relevant to the modification of the effective superradiant growth rate.}

Focusing on the $n=2$ level subspace, we construct an effective three-level Hamiltonian in the basis ${\ket{211},\ket{210},\ket{21-1}}$ for $\alpha\equiv\mu M\ll1$. The level mixing is organized by symmetry: the disk enters through source multipoles, namely projections onto $Y_{2m_d}$, and the selection rule $m_d=m'-m$ fixes which sublevels couple. Axisymmetric disks give only diagonal shifts; equatorially symmetric but nonaxisymmetric structures activate the $\Delta m=\pm2$ channel, mixing $\ket{211}\leftrightarrow\ket{21-1}$ while leaving $\ket{210}$ decoupled; and breaking equatorial-reflection symmetry opens $\Delta m=\pm1$ couplings that involve $\ket{210}$ and yield full three-level mixing. This symmetry-based classification is the central organizing principle of this work.

We illustrate these channels with two representative disk configurations. The first is a transient equatorial $m=2$ spiral density-wave packet with a Gaussian envelope, which reduces the problem to a driven two-level system and can suppress superradiance by transferring population into the decaying mode. The second is a quasi-static warped disk motivated by the Bardeen--Petterson scenario~\cite{Bardeen:1975zz,10.1093/mnras/202.4.1181}, whose static three-level mixing can open narrow ``growth gaps'' near accidental near-degeneracies and can either suppress or enhance the effective growth rate depending on the initial-state composition.

The remainder of the paper is organized as follows. In Sec.~\ref{Gravitational Atoms and Multi-Level Systems} we review the gravitational-atom framework and introduce the effective multi-level description used in this work. In Sec.~\ref{Accretion Disks as Environmental Perturbations} we establish the timescale hierarchy, formulate the disk perturbation in a spherical-harmonic expansion, and derive the symmetry-based selection rules. In Sec.~\ref{Representative Accretion-Disk Models} we present two representative disk models, equatorial spiral waves and static warps, and quantify their impact on level mixing and the effective growth rate. Sec.~\ref{sec:conclusion} is the conclusion and discussion. Technical details and auxiliary derivations are collected in the Appendices. Throughout this work, we use units $G=c=\hbar=1$.

\section{Gravitational Atoms and Multi-Level Systems}
\label{Gravitational Atoms and Multi-Level Systems}

 \subsection{Scalar Bound States and Superradiant Growth in Kerr Spacetime}
 
Consider a scalar field $\Phi$ of mass $\mu$ in Kerr spacetime, satisfying the Klein–Gordon equation
    \begin{equation}
    \left(g^{\mu\nu}\nabla_\mu\nabla_\nu-\mu^2\right)\Phi=0,
        \label{KG-eq}
    \end{equation}
where the Kerr metric $g_{\mu\nu}$ in Boyer–Lindquist coordinates is given by
\begin{equation}
        \d s^{2}=-\left(1-\frac{2Mr}{\Sigma}\right)\d t^2-\frac{4Mar\sin^2\theta}{\Sigma}\d t\d\phi+\frac{\Sigma}{\Delta}\d r^2+\Sigma \d\theta^2+\left(r^2+a^2+\frac{2Ma^2r\sin^2\theta}{\Sigma}\right)\sin^2\theta \d\phi^2
\end{equation}
with $\Sigma\equiv r^2+a^2\cos^2\theta$, $\Delta\equiv r^2-2Mr+a^2$ and the spin parameter $a\equiv J/M$. The event horizon radius is $r_+=M+\sqrt{M^2-a^2}$ and the BH angular velocity is $\Omega_H=\frac{a}{2M r_+}$. The solutions of Eq.~\eqref{KG-eq} can be separated as~\cite{Brill:1972xj}
 \begin{equation}
    \Phi(t,\bm r)=\e^{-\i\omega t}\e^{\i m\phi}S_{\ell m}(\theta)R_{n\ell}(r),
    \label{scalar_solution}
\end{equation}
Due to the presence of the BH horizon, the eigenvalue $\omega$ of Eq.~\eqref{scalar_solution} yields a complex eigenfrequency
    \begin{equation}
        \omega=\omega_R+\i\Gamma,
    \end{equation} 
where the real part $\omega_R$ corresponds to the gravitational atom spectrum and the imaginary part $\Gamma$ denotes the superradiant growth rate. Here we define $\omega_{R}\equiv\omega_{n \ell m}$ and $\Gamma\equiv\Gamma_{n\ell m}$. Explicit expressions for these quantities are given by~\cite{Detweiler:1980uk,Baumann:2018vus}
\begin{align}
    &\Gamma_{n\ell m}=\mu r_+ C_{n\ell m}(m\Omega_H-\omega)\alpha^{4\ell+4},
    \label{growth_rate}\\
    &\omega_{n\ell m}\simeq\mu\left(1-\frac{\alpha^2}{2n^2}-\frac{\alpha^4}{8n^4}+\frac{(2\ell-3n+1)\alpha^4}{n^4(\ell+1/2)}+\frac{2\tilde am\alpha^5}{n^3\ell(\ell+1/2)(\ell+1)}\right),
    \label{omegaR}
\end{align}
with 
\begin{align}
    C_{n\ell m}&=\frac{2^{4\ell+1}(2\ell+n+1)!}{(\ell+n+1)^{2\ell+4}n!}\left[\frac{\ell!}{(2\ell)!(2\ell+1)!}\right]^2\times\prod_{j=1}^{\ell}\left[j^2\left(1-a^2/M^2\right)+r_+^2\left(m\Omega_H-\omega\right)^2\right],
\end{align}
in the weak-coupling limit $\alpha\equiv\mu M\ll1$. Superradiance requires $\Gamma>0$, i.e. $m\Omega_H>\omega$. As the process proceeds, the occupation number of the growing state increases exponentially, and the boson cloud continuously extracts spin energy from the BH, reducing $\Omega_H$. When the system evolves to the critical condition $m\Omega_H\simeq\omega$, the growth rate $\Gamma\propto(m\Omega_H-\omega)$ vanishes, halting superradiance. At this stage, the boson cloud ceases to grow and the system reaches saturation, often referred to as the steady state of the gravitational atom, where the BH spin-down and boson cloud energy storage achieve dynamic equilibrium~\cite{Yoshino:2012kn}.

\subsection{Three-Level Truncation and Perturbative Mixing of Gravitational-Atom States}

\begin{figure}[h]
    \centering
    \includegraphics[width=0.5\textwidth]{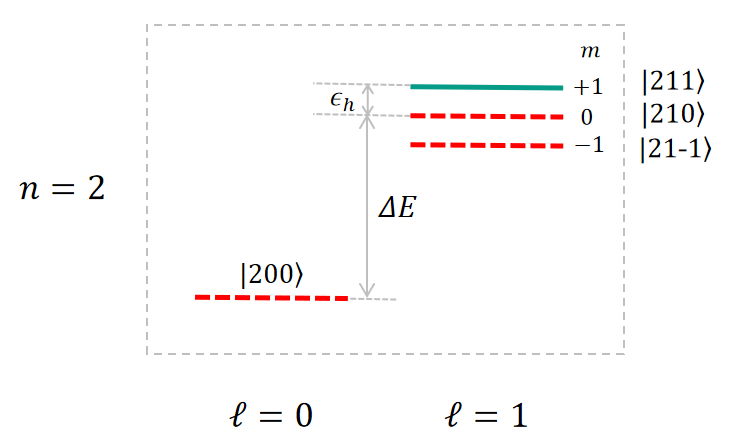}
    \caption{Partial energy levels of the gravitational atom. The states $\ket{211}$, $\ket{210}$, and $\ket{21-1}$ are shown together with the nearby $\ket{200}$ level. Green solid lines denote growing states, red dashed lines denote decaying states. $\Delta E$ is the energy difference between $\ket{210}$ and $\ket{200}$, while $\epsilon_h$ denotes the splitting between adjacent magnetic quantum numbers $m$ within the $\ket{21m}$ subspace.}
\end{figure}

We now restrict the dynamics to the nearly degenerate magnetic sublevels
$\{\ket{211},\ket{210},\ket{21-1}\}$ and study their mixing induced by the
accretion-disk perturbation. The full Hilbert space is decomposed as
\begin{equation}
    \mathcal H=P\oplus Q,
    \qquad
    P=\mathrm{span}\{\ket{211},\ket{210},\ket{21-1}\},
\end{equation}
where $Q$ contains all other bound and continuum states. Accordingly, the cloud state can be truncated as the $\ket{21m}$ triplet in the subspace $P$
\begin{equation}
    \ket{\psi_c}
    =
    c_g\ket{211}
    +
    c_d^{(1)}\ket{210}
    +
    c_d^{(2)}\ket{21-1}.
\end{equation}
Since the disk perturbation mainly redistributes probability weight between
growing and decaying levels, the effective growth rate is
\begin{equation}
    \Gamma_{\mathrm{eff}}
    =
    |c_g|^2\Gamma_{211}
    +
    |c_d^{(1)}|^2\Gamma_{210}
    +
    |c_d^{(2)}|^2\Gamma_{21-1},
\end{equation}
as derived in Appendix~\ref{A}. 

\color{blue}

However, to justify such a restriction, two important issues must be addressed. The first is whether the $\ket{21m}$ subspace can be consistently treated as a complete and approximately Hermitian sector. Strictly speaking, imposing the ingoing boundary condition at the horizon renders the scalar problem formally non-Hermitian. Nevertheless, in the weak-coupling regime where $\alpha\simeq0.03$--$0.1$, the imaginary widths $\Gamma_{211}\simeq \mu r_+ C_{211}(\Omega_H-\omega_R)\alpha^8$ are severely suppressed compared to the real hyperfine level splittings $\epsilon_h \equiv \omega_{211}-\omega_{210} \simeq \frac{1}{12}\mu\tilde a\alpha^5$, yielding the ratio
\begin{align}
\frac{\Gamma_{211}}{\epsilon_h}\sim 10^{-8}\text{--}10^{-6}.
\end{align}
Because this non-Hermitian contribution introduces only negligible corrections to the real quasi-bound wavefunctions, the $\ket{21m}$ triplet can be safely treated as an approximately complete and Hermitian subspace for the purpose of describing the gravitationally perturbed system within our approximation.

The second issue concerns the potential leakage of state population from this retained subspace $P$ to the external subspace $Q$. To quantify this effect, we define the leakage parameter $\eta_Q$ as the maximum state-mixing matrix element between the two sectors:
\begin{equation}
    \eta_Q
    \equiv
    \max_{a\in Q,\ i\in P}
    \left|
    \frac{\bra{a}V_{\rm disk}\ket{i}}
    {E_a-E_i}
    \right|.
\end{equation}
As demonstrated by the estimates in Appendix~\ref{app:eta}, $\eta_Q\ll1$ holds firmly across the entire parameter range of interest, ensuring that the coupling to external states is highly suppressed.

Within this context, the potential mixing with the nearby $\ket{200}$ state warrants special attention. As indicated by Eq.~\eqref{growth_rate} of the revised manuscript, the $\ell=m=0$ states are more strongly damped since the decay rate scales as $\Gamma_{n\ell m}\propto \alpha^{4\ell+4}$. Given its proximity in energy, the $\ket{200}$ state represents the most critical potential mixing partner, whose inclusion would further suppress the effective growth rate due to its rapid damping. However, this channel is strictly forbidden by angular selection rules. The leading gravitational multipole induced by the disk is the quadrupolar component with $\ell_d=2$. For an initial gravitational-atom state with $\ell=1$, the quadrupole selection rules dictate that
\begin{equation}
    |\ell-\ell_d|\leq \ell'\leq \ell+\ell_d.
\end{equation}
With $\ell=1$ and $\ell_d=2$, the allowed final orbital quantum numbers are strictly limited to $\ell'=1,3$, thereby forbidding transitions to $\ell'=0$. Consequently, despite being energetically nearby and possessing a faster decay rate, the $\ket{200}$ state does not directly couple to the retained $\{|211\rangle,|210\rangle,|21{-}1\rangle\}$ triplet under the dominant quadrupolar disk perturbation. This decoupling, alongside the condition $\eta_Q\ll1$, fully justifies the three-level truncation for both accretion-disk models considered in this work.

Complementary to the external leakage, we introduce the internal mixing parameter
\begin{equation}
    \eta_P
    \equiv
    \max_{i\neq j\in P}
    \left|
    \frac{\bra{i}V_{\rm disk}\ket{j}}
    {E_i-E_j}
    \right|
    =\max_{m\neq m^\prime}\left|\frac{H_{m,m^\prime}}{\epsilon_h}\right|,
\end{equation}
which measures the strength of the disk-induced mixing strictly within the subspace $P$. Within this retained sector, the unperturbed Hamiltonian takes the diagonal form
\begin{equation}
    H_{0,P}=
    \begin{pmatrix}
        E+\epsilon_h & 0 & 0\\
        0 & E & 0\\
        0 & 0 & E-\epsilon_h
    \end{pmatrix},
    \qquad
    E\equiv\omega_{210}
    =
    -\mu\left(
        \frac{1}{8}\alpha^2
        +
        \frac{7}{128}\alpha^4
    \right),
\end{equation}
where the distinct levels are designated as $E_1\equiv E+\epsilon_h$, $E_2\equiv E$, and $E_3\equiv E-\epsilon_h$. Physically, while $\eta_Q\ll1$ validates the mathematical consistency of the three-level truncation, $\eta_P\ll1$ serves as the additional prerequisite for applying non-degenerate perturbation theory inside $P$. In regimes where $\eta_P\sim1$ or larger, the truncation remains valid as long as $\eta_Q\ll1$, but the precise dynamics must be resolved either by diagonalizing the near-degenerate block or by directly solving the projected Schr\"odinger equation.

\color{black}

\color{black}

\section{Accretion Disks as Environmental Perturbations}
\label{Accretion Disks as Environmental Perturbations}

Under the strong gravity of the BH, captured matter drifts inward while transporting angular momentum outward, forming an accretion disk~\cite{Shakura1973,1974MNRAS.168..603L,1981ARA&A..19..137P}. Only electrically neutral, non-magnetized disks are considered, so that the disk affects the gravitational atom solely through gravity. For simplicity, the disk is further assumed to be geometrically thin, lying on the equatorial plane or tilted by a small inclination angle. In this limit, the disk is described by a surface density $\Sigma(t,r,\phi)$~\cite{2002apa..book.....F}. In this work, we aim to treat the disk's gravitational influence as a weak perturbation, rendering the following foundational discussions essential.

\subsection{Working Domain of $\alpha$ from Timescale and Multipole Constraints}

To rigorously model the accretion disk as a weak external perturbation on a fixed Kerr background, two independent conditions must be satisfied. First, the superradiant growth of the boson cloud must be much faster than the accretion-induced drift of the BH's mass and spin. Second, the disk potential within the cloud region must admit a rapidly convergent low-multipole expansion. Below, we evaluate these timescale and geometric constraints in sequence to determine the valid working domain for the gravitational fine structure constant $\alpha$.

\subsubsection{Timescale Hierarchy and Negligible Accretion-Driven Evolution}

Before modeling the disk gravity as an external perturbation, we first justify
that the secular evolution of the Kerr parameters induced by accretion is
negligible during the \emph{linear} superradiant growth stage~\cite{Arvanitaki:2010sy}. The physical
question is simple: does accretion change the horizon angular velocity fast
enough to noticeably move the system away from (or toward) the superradiant
condition $m\Omega_H>\omega_R$ while the cloud is growing? This is answered by
comparing two timescales,
\begin{equation}
    \tau_{\mathrm{sr}} \equiv \Gamma_{211}^{-1},
    \qquad
    \tau_{\Omega} \equiv \frac{\Omega_H}{\left|\dot{\Omega}_H\right|}.
\end{equation}
Here $\tau_{\mathrm{sr}}$ characterizes how rapidly the dominant level
$\ket{211}$ amplifies, while $\tau_{\Omega}$ characterizes how rapidly accretion
drifts $\Omega_H$. If $\tau_{\mathrm{sr}}\ll\tau_{\Omega}$, then $M$ and $\tilde a$
may be treated as constants over the growth epoch, and the disk can be regarded
as an external perturbation acting on an effectively fixed Kerr background. In
practice, we impose a conservative separation $\tau_{\mathrm{sr}}\le
\epsilon\,\tau_{\Omega}$ with $\epsilon\ll 1$.

We begin with $\tau_{\Omega}$, because it is set primarily by the
\emph{time-averaged} accretion strength. The Eddington luminosity
$L_{\mathrm{Edd}}$ corresponds to the balance between gravity and radiation
pressure on ionized gas (via Thomson scattering). Assuming a radiative
efficiency $L=\eta \dot M $~\cite{2002apa..book.....F,novikov1973astrophysics}, it defines the Eddington accretion rate
\begin{equation}
    \dot M_{\mathrm{Edd}} \equiv \frac{L_{\mathrm{Edd}}}{\eta }.
\end{equation}
It is then convenient to parametrize the long-term inflow as
\begin{equation}
    \langle \dot M\rangle \simeq D\, f_{\mathrm{Edd}}\, \dot M_{\mathrm{Edd}},
\end{equation}
where $f_{\mathrm{Edd}}\equiv L/L_{\mathrm{Edd}}$ measures the accretion strength
relative to the Eddington limit and $D\le 1$ accounts for the intermittency (duty
cycle)~\cite{2018ApJ...857...53C,2008MNRAS.388..625S}. Defining the Salpeter time
\begin{equation}
    \tau_S \equiv \frac{M}{\dot M_{\mathrm{Edd}}}
    \simeq 4.5\times 10^{7}\,\mathrm{yr},
\end{equation}
one has $M/\langle \dot M\rangle=\tau_S/(D f_{\mathrm{Edd}})$, so the accretion
timescale is essentially controlled by the single combination $D f_{\mathrm{Edd}}$.
Differentiating $\Omega_H(M,\tilde a)$ and using $\dot J=l_{\mathrm{isco}}\dot M$, where $l_{\mathrm{isco}}$ is the specific angular momentum at the innermost stable circular orbit (ISCO) (see Appendix~\ref{B}).
\begin{equation}
    \tau_{\Omega}
    \simeq
    g(\tilde a)\,\frac{\tau_S}{D f_{\mathrm{Edd}}},
    \qquad
    g(\tilde a)\equiv
    \left|
    \frac{\tilde a\sqrt{1-\tilde a^2}}
    {\tilde l_{\mathrm{isco}}-\tilde a\left(2-\sqrt{1-\tilde a^2}\right)}
    \right|,
    \label{eq:tauOmega_g}
\end{equation}
where $g(\tilde a)=\mathcal O(1)$ encodes the order-unity dependence on the spin and on the specific angular momentum at the ISCO.

We next estimate $\tau_{\mathrm{sr}}$ for the dominant $\ket{211}$ level. In the
weak-coupling regime, Eq.~\eqref{growth_rate} yields
\begin{equation}
    \Gamma_{211}
    =
    \mu r_+\,C_{211}\,(\Omega_H-\omega_R)\,\alpha^8,
    \qquad
    \tau_{\mathrm{sr}}=\Gamma_{211}^{-1}.
    \label{eq:Gamma211_timescale}
\end{equation}
Using $\mu=\alpha/M$ and defining the dimensionless detuning
\begin{equation}
    \Omega_H-\omega_R \equiv \frac{\tilde\Delta(\alpha)}{M},
    \qquad
    \tilde\Delta(\alpha)\equiv M(\Omega_H-\omega_R),
\end{equation}
one obtains the scaling form
\begin{equation}
    \tau_{\mathrm{sr}}
    =
    \frac{M}{(r_+/M)\,C_{211}\,\tilde\Delta(\alpha)}\,\alpha^{-9},
    \label{eq:tausralpha9}
\end{equation}
up to slowly varying prefactors. A key point for quantitative estimates is that
$(r_+/M)C_{211}$ is \emph{not} generically order unity; for $\tilde a=\mathcal
O(1)$ one typically finds $C_{211}\sim 10^{-3}$, so replacing this prefactor by
an $\mathcal O(1)$ constant can shift $\tau_{\mathrm{sr}}$ by orders of
magnitude.

Combining Eqs.~\eqref{eq:tauOmega_g} and \eqref{eq:tausralpha9} with the
conservative requirement $\tau_{\mathrm{sr}}\le \epsilon\,\tau_{\Omega}$ yields
an implicit lower bound on $\alpha$,
\begin{equation}
    \alpha_{\min}
    \simeq
    \left[
    \frac{M\,D f_{\mathrm{Edd}}}{\epsilon\,g(\tilde a)\,\tau_S}\cdot
    \frac{1}{(r_+/M)\,C_{211}\,\tilde\Delta(\alpha_{\min})}
    \right]^{1/9}
    \label{eq:alpha_min_general}
\end{equation}
which makes the logic transparent: the \emph{worst case} for neglecting accretion
corresponds to the largest time-averaged inflow $D f_{\mathrm{Edd}}$ (shortest
$\tau_\Omega$), while intermittent accretion ($D\ll 1$) only lengthens
$\tau_\Omega$ and relaxes the bound as $\alpha_{\min}\propto (D f_{\mathrm{Edd}})^{1/9}$.

Finally, to obtain a concrete numerical estimate, we adopt a conservative
benchmark with persistent accretion ($D\simeq 1$) and a sub-Eddington strength
$f_{\mathrm{Edd}}=0.1$, representative of luminous but not near-Eddington disks
(thereby avoiding strong dependence on radiative feedback while still maximizing
secular drift). For a rapidly spinning stellar-mass Kerr BH with $M=10M_\odot$
and $\tilde a=0.9$, we impose the conservative separation
$\tau_{\mathrm{sr}}\lesssim 0.01\,\tau_{\Omega}$. Using the scaling in Eq.~\eqref{eq:tausralpha9} together with
$\tau_{\Omega}\simeq g(\tilde a)\tau_S/(D f_{\mathrm{Edd}})$, and adopting
representative values $g(\tilde a)=\mathcal O(1)$,
$C_{211}\sim 10^{-3}$ and $\tilde\Delta(\alpha)\sim\mathcal O(0.1)$ in the
regime of interest, we obtain a lower bound
\begin{equation}
     \alpha\gtrsim0.02,
    \label{eq:alpha_min}
\end{equation}
for which the accretion-driven evolution of $M$ and $\tilde a$ is slow compared
with the linear superradiant growth.

Physically, the lower bound arises because in the weak-coupling regime the
superradiant rate is strongly suppressed at small $\alpha$ (for $\ell=1$,
$\Gamma_{211}\sim \alpha^9$), reflecting the exponentially small near-horizon
wave function amplitude when the cloud radius scales as $r_c\sim M/\alpha^2$.
For sufficiently small $\alpha$, the growth becomes so slow that secular
accretion-driven drift of $\Omega_H$ can no longer be neglected during the
linear stage. We stress that this estimation is primarily
controlled by the time-averaged inflow $D f_{\mathrm{Edd}}$ and scales as
$\alpha_{\min}\propto (D f_{\mathrm{Edd}})^{1/9}$; for intermittent accretion
($D\ll 1$) the bound becomes correspondingly weaker. In the remainder of this
work we therefore treat $M$ and $\tilde a$ as constants during the linear-growth
stage and evaluate the level-dependent growth/decay rates $\Gamma$ using
Eq.~\eqref{growth_rate}.

\begin{figure}[h]
    \centering
    \includegraphics[width=0.7\linewidth]{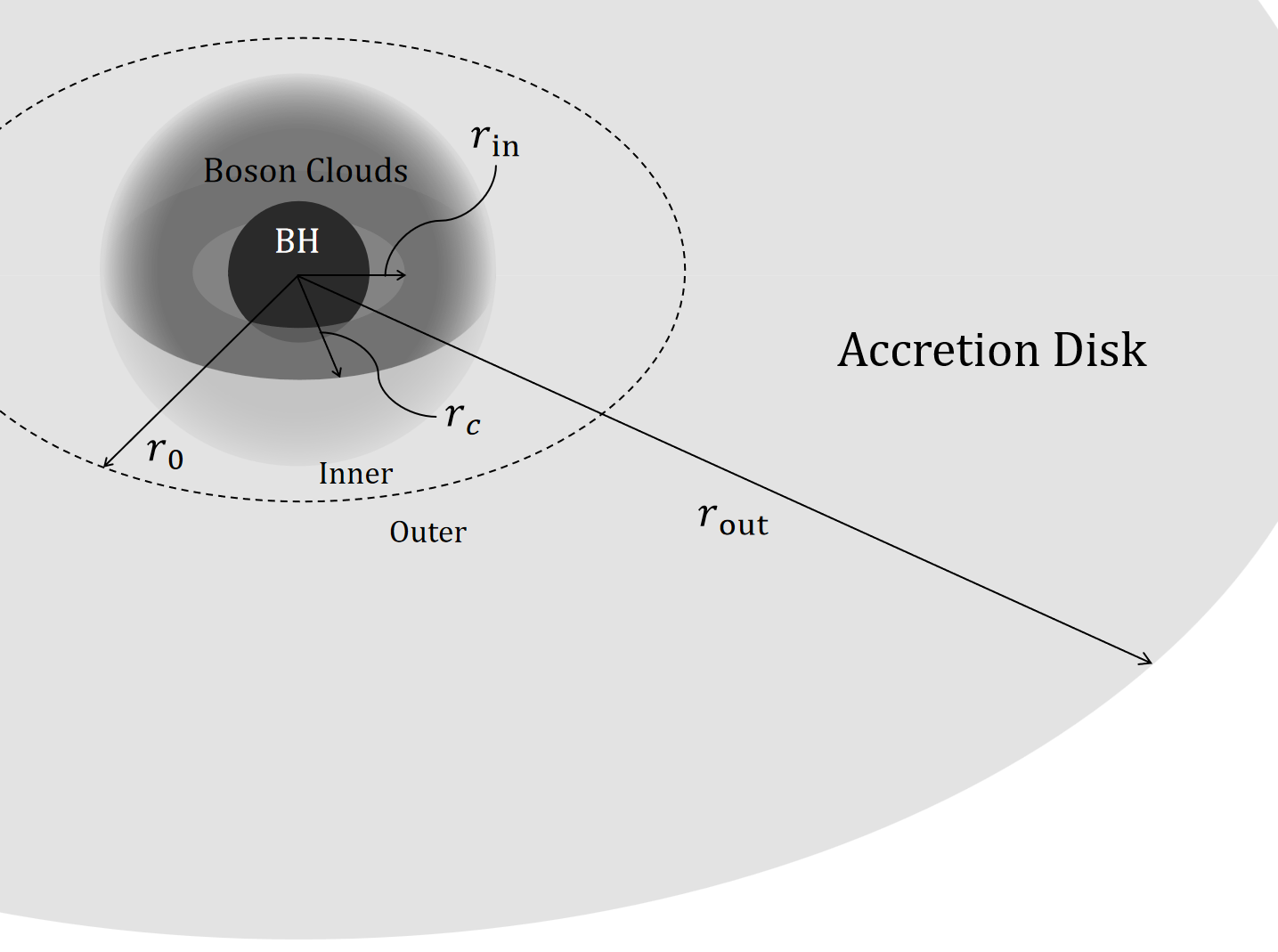}
    \caption{Schematic configuration of the BH, the boson cloud, and a geometrically thin accretion disk. The cloud is characterized by a typical radius $r_c$, while the disk extends from the inner radius $r_{\rm in}$ to the outer radius $r_{\rm out}$. The hierarchy $r_{\rm in}<r_c<r_{\rm out}$ illustrates that the cloud typically resides well inside the bulk of the disk. {\color{blue}For example, for $\tilde a=0.9$ and $\alpha=0.1$, one has $r_{\rm in}=r_{\rm isco}\approx 2.32M$, $r_c\simeq 400M$, and a representative outer radius $r_{\rm out}\sim 10^5M$.}}
    \label{fig:relation}
\end{figure}

\subsubsection{Validity Domain for a Perturbative Multipole Treatment}
\label{sec:disk_validity}

We have shown above that, for $\alpha\gtrsim 0.02$ given by Eq.~\eqref{eq:alpha_min},
the superradiant growth of the dominant $\ket{211}$ level is fast compared with the
secular accretion-driven drift of the Kerr parameters, so that $M$ and $\tilde a$
may be treated as constants during the linear stage. We now impose the additional
requirement needed to model the disk gravity as an \emph{external} perturbation:
in the cloud region, the disk potential should admit a rapidly convergent spherical-harmonic
(multipole) expansion~\cite{Thorne:1980ru} that is dominated by low $\ell$~\cite{Baumann:2018vus}. Since the
addition theorem for $1/|\bm r-\bm r'|$ converges differently depending on whether the
field point lies inside or outside the source point, it is natural to split the disk at a
radius $r_0\gtrsim r_c$ into an ``inner'' part ($r_{\rm in}<r'<r_0$) and an ``outer'' part
($r_0<r'<r_{\rm out}$). The perturbative description is quantitatively controlled provided that the following two conditions are satisfied:
\begin{enumerate}
    \item the cloud predominantly feels an \emph{outer} tidal field (so that the outer-source expansion with convergence parameter $r/r'<1$ is valid throughout the cloud support). This condition is quantified by
    \begin{align}
        \epsilon_\Phi(r_0)\equiv
    \left|\frac{\Phi_{\mathrm{in}}(r_c;r_0)}{\Phi_{\mathrm{out}}(r_c;r_0)}\right|\ll 1
    \end{align}
    where
    \begin{equation}
    \Phi_{\mathrm{in}}(r_c;r_0)
    \sim
    \frac{2\pi}{r_c}
    \int_{r_{\mathrm{in}}}^{r_0}
    \Sigma(r'),r',dr',
    \qquad
    \Phi_{\mathrm{out}}(r_c;r_0)
    \sim
    2\pi
    \int_{r_0}^{r_{\mathrm{out}}}
    \Sigma(r'),dr' .
    \label{eq}
    \end{equation}
    Here the split radius $r_0$ is chosen slightly larger than the characteristic  radius of the cloud, $r_c$, to ensure rapid convergence within the cloud region.
    
    \item The disk material within the cloud does not significantly back-react on the cloud profile. This requirement is measured by 
    \begin{align}
        \epsilon_M(r_c)\equiv \frac{M_{\rm in}(r_c)}{M_c}\ll 1,
        \label{eq:eps_def_general_compact}
    \end{align}
    where the mass of the accretion disk inside the cloud $M_{\rm in}(r_c)=2\pi\int_{r_{\rm in}}^{r_c}\Sigma(r')\,r'\,dr'$ and the total mass stored in the scalar cloud $M_c$.
\end{enumerate}
{\color{blue}
In general, for a given disk model, the applicability of the multipole expansion should be assessed by requiring both $\epsilon_\Phi$ and $\epsilon_M$ to remain below a prescribed tolerance $\epsilon_{\rm tol}$,
\begin{equation}
\epsilon_\Phi(\alpha;p,x_{\rm in},x_{\rm out})<\epsilon_{\rm tol},
\qquad
\epsilon_M(\alpha;p,\Sigma_0M,M_c/M,x_{\rm in},x_{\rm out})<\epsilon_{\rm tol},
\end{equation}
and equivalently the threshold value $\alpha_{\rm multi}$ is defined as the infimum of the set of $\alpha$ values satisfying both conditions,
\begin{equation}
\alpha_{\rm multi}
=
\inf\left\{
\alpha \,\middle|\,
\epsilon_\Phi(\alpha)<\epsilon_{\rm tol},
\quad
\epsilon_M(\alpha)<\epsilon_{\rm tol}
\right\}.
\label{eq:alpha_multi_general}
\end{equation}
\color{black}
Here, we adopt an axisymmetric equatorial power-law disk
profile~\cite{Shakura1973,1974MNRAS.168..603L}
\begin{equation}
    \Sigma(r)=\Sigma_0\left(\frac{r}{M}\right)^{-p},
    \label{eq:Sigma_powerlaw_compact}
\end{equation}
and then the two control parameters become
\begin{equation}
\epsilon_\Phi
\sim
\frac{1-p}{4(2-p)},
\frac{4^{2-p}-(\alpha^2x_{\mathrm{in}})^{2-p}}
{(\alpha^2x_{\mathrm{out}})^{1-p}-4^{1-p}},
\quad
\epsilon_M
=
\frac{2\pi}{\mu_c(2-p)}
(\Sigma_0 M)
\left[4^{2-p}-(\alpha^2x_{\mathrm{in}})^{2-p}\right]
\alpha^{-2(2-p)},
\label{eq:epsPhi_epsM_powerlaw_compact}
\end{equation}
where $x_{\mathrm{in}}=r_{\mathrm{in}}/M$ and $x_{\mathrm{out}}=r_{\mathrm{out}}/M$.  In addition, we parametrize the cloud mass $M_c$ as $\mu_c=\frac{M_c}{M}$, with $\mu_c$ the cloud-to-BH mass ratio.

\color{blue}
Since more massive BHs, such as active galactic nuclei (AGNs) \cite{1993ARA&A..31..473A}, are embedded in more complicated environments, a complete treatment would require a detailed discussion of environmental effects beyond the accretion disk. In this work, we therefore focus on the relatively clean case of stellar-mass BHs with accretion disks, which we refer to as the stellar-mass accretion-disk.
For stellar-mass thin disks, the dimensionless surface-density normalization is very small~\cite{Abramowicz:2011xu,Abramowicz:1988sp}
\begin{align}
\Sigma_0M\simeq10^{-15},
\end{align}
which makes $\epsilon_M\ll \epsilon_\Phi$ in the parameter range considered here. Thus the multipole validity condition is mainly controlled by the potential-ratio parameter $\epsilon_\Phi$.
\color{black}
Using representative thin-disk parameters $x_{\mathrm{in}}=2.32$, $x_{\mathrm{out}}=10^{5}$~\cite{novikov1973astrophysics,2002apa..book.....F}, $p=3/4$, and $\mu_c=10^{-2}$~\cite{Brito:2015oca}, one finds $\epsilon_\Phi(\alpha=0.1)\simeq6.7\times10^{-2}$ and $\epsilon_\Phi(\alpha=0.03)\simeq1.7\times10^{-1}$. Requiring, for definiteness, $\epsilon_\Phi\lesssim0.2$ gives the conservative estimate
\begin{equation}
\alpha \gtrsim 0.03 .
\label{eq:alpha_min_multipole_compact}
\end{equation}
Combining this with the timescale requirement in Eq.~\eqref{eq:alpha_min}, we take the working domain to be
\begin{equation}
\alpha \gtrsim
\max\left[
0.02\ \text{(from timescales)},
0.03\ \text{(from disk multipoles)}
\right]
\simeq 0.03,
\label{eq:alpha_working_domain_compact}
\end{equation}
within which the disk self-gravity can be treated as a weak, external, low-multipole perturbation acting on the hydrogenic cloud basis.

\subsection{Spherical-Harmonic Expansion of the Disk Gravitational Perturbation}

Having established the working domain of $\alpha$, we next proceed to model the gravitational potential of the accretion disk and perform the spherical-harmonic expansion. Based on this expansion, we then systematically analyze the symmetry principles of the disk configuration.

\subsubsection{Relativistic correction estimate based on quadrupolar matrix elements}
\label{subsubsec:gr_correction_quadrupole}

Although the unperturbed scalar spectrum is computed in the Kerr background, we nevertheless adopt the Newtonian gravitational potential to model the disk-induced perturbation. We therefore estimate whether the omitted relativistic corrections to the disk potential can significantly modify the leading quadrupolar matrix elements. As shown below, for the smooth thin-disk profiles considered here, these corrections are parametrically suppressed and do not affect the selection rules or the qualitative mixing channels.

\color{blue}

It is useful to separate the accretion disk into an inner region and an outer region bounded by $r_1$. Unlike $r_0$ introduced in Sec.~\ref{sec:disk_validity}, which solely serves the purpose of the perturbative multipole treatment, $r_1$ divides the accretion disk into the following two distinct parts. The inner disk, being close to the BH, is strongly affected by relativistic frame-dragging effects, including the Lense--Thirring effect \cite{Mashhoon:1984rj}, the Bardeen--Petterson effect \cite{Bardeen:1975}, and geodetic precession \cite{deSitter:1916}. By contrast, the outer disk is less sensitive to these strong-field relativistic effects.

{We define near-BH inner-to-outer Hamiltonian ratio
\begin{equation}
\epsilon_H^{(2)}
\equiv
\left|
\frac{H_{\rm in}^{(2)}}{H_{\rm out}^{(2)}}
\right| ,
\end{equation}
where, using the $2p$ radial expectation values,
\begin{equation}
H_{\rm in}^{(2)}
\sim
\mu\langle r^{-3}\rangle_{2p}
\int_{r_{\rm in}}^{r_1}\Sigma(r')r'^3\d r',
\qquad
H_{\rm out}^{(2)}
\sim
\mu\langle r^2\rangle_{2p}
\int_{r_1}^{r_{\rm out}}\Sigma(r')r'^{-2}\d r' .
\end{equation}
Usually, taking $r_1 \simeq 10M$ and adopting the power-law profile $\Sigma(r) = \Sigma_0 (r/M)^{-p}$, this gives
\begin{equation}
\epsilon_H^{(2)}
\sim
\frac{\alpha^{10}}{720}
\frac{1+p}{4-p}
\frac{
10^{4-p}-x_{\rm in}^{4-p}
}{
10^{-1-p}-x_{\rm out}^{-1-p}
} .
\label{eq:epsilon_H2_powerlaw}
\end{equation}
Taking a conservative value of $\alpha \simeq 0.05$ and adopting the same parameters as before, we obtain
\begin{equation}
    \epsilon_H^{(2)} \simeq 7.2 \times 10^{-12} \ll 1,
\end{equation}
which implies that the contribution of the inner disk to the relativistic corrections is strongly suppressed.

For the outer region of the disk, which is slow-moving and far from the BH, the leading relativistic corrections may be estimated locally by
\begin{equation}
    \frac{\delta H_{\rm GR}}{H_{\rm N}}
    \sim
    \mathcal O\!\left(\frac{M}{r'}\right)
    +\mathcal O(v_{\rm disk}^2)
    +\mathcal O(v_{\rm cloud}^2) .
\end{equation}
where the Keplerian orbital correction $v_{\rm disk}^2\sim M/r'$, and $v_{\rm cloud}^2\sim\alpha^2$ are subleading for the present first-order perturbative estimate. Therefore the relativistic correction to the quadrupolar matrix element can be obtained by averaging
$q(r')\sim M/r'$ with the outer quadrupolar weights:}
\begin{equation}
\eta_{\rm GR}^{(2)}
\equiv
\left|
\frac{\delta H_{\rm GR}^{(2)}}{H_{\rm N}^{(2)}}
\right|
\sim
\frac{
\left(\delta H^{(2)}_{\rm out}/H_{\rm out}^{(2)}\right)
+
\epsilon_H^{(2)}
\left(\delta H^{(2)}_{\rm in}/H^{(2)}_{\rm in}\right)
}{1+\epsilon_H^{(2)}}
\sim\frac{\delta H^{(2)}_{\rm out}}{H_{\rm out}^{(2)}}.
\end{equation}
For the power-law disk one finds
\begin{equation}
\frac{\delta H^{(2)}_{\rm out}}{H_{\rm out}^{(2)}}\sim\frac{\int_{r_1}^{r_{\rm out}}q(r')\Sigma(r')r'^{-2}\d r'}{\int_{r_1}^{r_{\rm out}}\Sigma(r')r'^{-2}\d r'}\simeq\frac{p+1}{p+2}\frac{M}{r_1}
\end{equation}
For the representative value $p=3/4$, this becomes
\begin{equation}
\eta_{\rm GR}^{(2)}
\simeq
0.064 .
\end{equation}

Therefore, in the weak-coupling regime considered here, the relativistic correction to the disk-induced quadrupolar matrix element is remarkably small. It can shift the numerical value of the matrix elements, and hence the precise location of narrow resonant or near-degenerate features, but it does not change the leading selection rules, mixing channels, or qualitative conclusions. A fully relativistic treatment of the disk perturbation would be necessary only for configurations in which the disk quadrupole is dominated by the innermost relativistic region, or for thick, magnetized, massive, or strongly self-gravitating disks.

\color{black}
\subsubsection{Quadrupolar Perturbation}
\label{sec:disk_quadrupole}

As discussed in the previous subsection, the relativistic corrections to the
disk-induced quadrupolar matrix elements are parametrically suppressed in the
smooth thin-disk regime considered here. We therefore keep the leading
gravitoelectric contribution of the disk self-gravity and model it by the
Newtonian potential in a freely falling frame centered on the BH--cloud center
of mass,
\begin{equation}
    U(t,\bm r)
    =
    -\int
    \frac{\rho(t,\bm r')}{|\bm r-\bm r'|}
    \,\mathrm{d}V' .
    \label{N-potential}
\end{equation}
where $\rho(t,\bm r')$ is the mass volume density. The spherical-harmonic addition theorem,
\begin{equation}
    \frac{1}{|\bm r-\bm r'|}
    =
    \sum_{\ell=0}^{\infty}\sum_{m=-\ell}^{\ell}
    \frac{4\pi}{2\ell+1}\,
    \frac{r^{\ell}}{{r'}^{\ell+1}}\,
    Y_{\ell m}(\theta,\phi)\,Y_{\ell m}^\ast(\theta',\phi'),
\end{equation}
then yields a multipole expansion of Eq.~\eqref{N-potential}. The perturbing potential energy experienced by the boson cloud is
\begin{equation}
    V(t,\bm r)=\mu U(t,\bm r)
    =
    -\sum_{\ell_d=0}^{\infty}\sum_{m_d=-\ell_d}^{\ell_d}
    \int
    \frac{4\pi\mu\rho(t,\bm r')}{2\ell_d+1}\,
    \frac{r^{\ell_d}}{{r'}^{\ell_d+1}}\,
    Y_{\ell_d m_d}(\theta,\phi)\,
    Y_{\ell_d m_d}^\ast(\theta',\phi')\,\mathrm{d}V'.
    \label{potential_energy}
\end{equation}
\color{blue}
The monopole contribution produces a common energy shift in the freely falling frame centered on the BH--cloud system, and therefore does not induce transitions and affect the effective growth rate through state mixing. The dipole contribution corresponds to a uniform gravitational field acting on the BH--cloud system and vanishes in the same freely falling frame. The quadrupole is therefore the lowest multipole producing a state-mixing effect within the $\ket{21m}$ subspace, which is directly relevant to the modification of the effective superradiant
growth rate 
\color{black}
\begin{equation}
    V_2(t,\bm r)
    =
    -\frac{4\pi\mu}{5}
    \sum_{m_d=-2}^{2}
    \int
    \rho(t,\bm r')\,
    \frac{r^{2}}{{r'}^{3}}\,
    Y_{2 m_d}(\theta,\phi)\,
    Y_{2 m_d}^\ast(\theta',\phi')\,\mathrm{d}V'.
\end{equation}
In the unperturbed basis, the perturbation matrix is
\begin{equation}
    H_{m,m'}=\braket{21m|V_2|21m'},
    \label{Hamiltonian}
\end{equation}
where the hydrogenic wavefunctions follow from Eq.~\eqref{scalar_solution},
\begin{equation}
    \braket{\bm r|n\ell m}\simeq e^{i\varepsilon_n t}\,R_{n\ell}(r)\,Y_{\ell m}(\theta,\phi).
\end{equation}
This perturbation couples different levels of the gravitational atom and induces level mixing.

\subsubsection{Symmetry Considerations and Selection Rules}
\label{Sec:symmetry}

The disk--induced level mixing is fixed once the quadrupolar ($\ell_d=2$) part of
the perturbing potential is projected onto the unperturbed $\{\ket{21m}\}$
subspace. Starting from the $\ell_d=2$ term of the spherical-harmonic addition
theorem, the quadrupolar contribution to the potential energy can be organized
into ``source multipoles'' by defining
\begin{equation}
    I_d^{(m_d)}(t)
    \equiv
    \int
    \frac{\rho(t,\bm r')}{{r'}^{3}}
    Y_{2 m_d}^\ast(\theta',\phi')\,\mathrm{d}V',
    \label{eq:Id_def}
\end{equation}
so that the perturbation experienced by the cloud takes the compact form
\begin{equation}
    V_2(t,\bm r)
    =
    -\frac{4\pi\mu}{5}\,r^2
    \sum_{m_d=-2}^{2}
    I_d^{(m_d)}(t)\,
    Y_{2m_d}(\theta,\phi),
    \label{eq:V2_Id}
\end{equation}
which makes explicit that all disk geometry and time dependence enter only
through $I_d^{(m_d)}(t)$. The perturbation matrix in the unperturbed basis is
\begin{equation}
    H_{m,m'}(t)
    \equiv
    \braket{21m|V_2(t,\bm r)|21m'}
    =
    \int \d^3r\,
    \psi_{21m}^\ast(\bm r)\,
    V_2(t,\bm r)\,
    \psi_{21m'}(\bm r),
    \label{eq:H_def}
\end{equation}
with $\psi_{21m}(\bm r)=e^{i\varepsilon_2t}R_{21}(r)Y_{1m}(\theta,\phi)$ and inserting Eq.~\eqref{eq:V2_Id} yields a fully factorized expression,
\begin{equation}
    H_{m,m'}(t)
    =
    -\frac{4\pi\mu}{5}
    \sum_{m_d=-2}^{2}
    I_d^{(m_d)}(t)\,
    \Bigl[\int_0^\infty \d r\,r^4 R_{21}^2(r)\Bigr]\,
    \Bigl[\int \d\Omega\,
    Y_{2m_d}(\Omega)\,Y_{1m}(\Omega)\,Y_{1m'}^\ast(\Omega)\Bigr].
    \label{eq:H_factorized}
\end{equation}
The first bracket in Eq.~\eqref{eq:H_factorized} is purely radial. For the
hydrogenic $2p$ wavefunction one finds
\begin{equation}
    \int_0^\infty \d r\,r^4 R_{21}^2(r)=30a^2,
    \qquad
    a\equiv(\mu\alpha)^{-1}=\frac{M}{\alpha^2},
\end{equation}
so that $\mu a^2=M/\alpha^3$. The second bracket is the angular integral
\begin{equation}
    I_\Omega(m,m';m_d)\equiv
    \int \d\Omega\,
    Y_{2m_d}(\Omega)\,Y_{1m}(\Omega)\,Y_{1m'}^\ast(\Omega),
    \label{eq:Iomega_def}
\end{equation}
which enforces the usual azimuthal selection rule. Using
$Y_{\ell m}(\theta,\phi)\propto e^{im\phi}$, the $\phi$ integral implies that
the integrand is nonzero only when the net phase is $\phi$-independent, i.e.
\begin{equation}
    m_d + m - m' = 0
    \qquad\Longrightarrow\qquad
    m_d = m' - m.
    \label{eq:md_selection}
\end{equation}
Therefore only the disk multipole with $m_d=m'-m$ can contribute to the matrix
element $H_{m,m'}$. Substituting the radial result and the definition
Eq.~\eqref{eq:Iomega_def} back into Eq.~\eqref{eq:H_factorized} gives
\begin{equation}
    H_{m,m'}(t)
    =
    -\frac{24\pi M}{\alpha^3}
    \sum_{m_d=-2}^{2}
    I_d^{(m_d)}(t)\,I_\Omega(m,m';m_d),
    \label{eq:H_mm_final}
\end{equation}
which makes the logic transparent: the disk determines which $m_d$ channels are
present via $I_d^{(m_d)}(t)$, while angular-momentum addition fixes which
$\ket{21m}$ sublevels can mix through Eq.~\eqref{eq:md_selection}.

Consequently, the symmetries of the disk determine which $m_d$ channels are
present, and Eq.~\eqref{eq:md_selection} then specifies which $\ket{21m}$
sublevels can mix. The key point is that $I_d^{(m_d)}$ is just the projection of
the disk density onto the quadrupolar spherical harmonic $Y_{2m_d}$,
\begin{equation}
    I_d^{(m_d)}(t)=\int \frac{\rho(t,\bm r')}{r'^3}\,Y_{2m_d}^\ast(\theta',\phi')\,\d V'.
    \label{Id}
\end{equation}
Therefore any symmetry that makes the disk ``orthogonal'' to a given $Y_{2m_d}$
forces that coefficient to vanish. If the disk is \emph{axisymmetric} ($\rho$ independent of $\phi'$)~\cite{Shakura1973,1974MNRAS.168..603L}, then only the
$m_d=0$ Fourier harmonic exists and all $m_d\neq 0$ projections vanish:
\begin{equation}
    I_d^{(m_d\neq 0)}=0.
\end{equation}
If the disk breaks axial symmetry but remains \emph{mirror-symmetric about the
equatorial plane}~\cite{Papaloizou:1995ss,1983PASJ...35..249K}, then the $\ell=2$ harmonics with $m_d=\pm1$ are forbidden because they are \emph{odd} across the midplane: \begin{equation} Y_{2,\pm1}\ \propto\ \sin\theta'\cos\theta'\,e^{\pm i\phi'} \quad\stackrel{\theta'\to\pi-\theta'}{\longrightarrow}\quad -\,Y_{2,\pm1}, \end{equation} so the contributions from above and below the plane cancel in the integral and \begin{equation} I_d^{(\pm1)}=0\qquad (\text{mirror-symmetric disk}). \end{equation} By contrast $Y_{2,0}$ and $Y_{2,\pm2}$ are even across the midplane, so $m_d=0,\pm2$ are allowed whenever the disk has the corresponding azimuthal structure. In short: \emph{azimuthal symmetry} kills $m_d\neq 0$, and \emph{mirror-symmetric} reflection kills the ``odd'' $m_d=\pm1$ quadrupole; getting $m_d=\pm1$ requires
breaking equatorial reflection.

For the $\{\ket{211},\ket{210},\ket{21-1}\}$ subspace, three cases are particularly relevant:
\begin{itemize}
    \item \textit{Axisymmetric disk} ($\partial_{\phi'}\Sigma=0$): the $\phi'$ integral in Eq.~\eqref{Id} vanishes unless $m_d=0$, and Eq.~\eqref{eq:md_selection} forces $m=m'$. The perturbation is therefore purely diagonal and induces no level mixing.
    \item \textit{Equatorial but nonaxisymmetric perturbations}: $m_d=\pm2$ components are generated, opening the $\Delta m=\pm2$ channel that mixes $\ket{211}\leftrightarrow\ket{21-1}$. If the mass distribution remains symmetric under reflection about the equatorial plane, $V_2$ is even parity and matrix elements connecting the odd-parity state $\ket{210}$ to the even-parity states $\ket{211}$ and $\ket{21-1}$ vanish, so the dynamics reduces to an effective two-level system.
    \item \textit{Out-of-equatorial geometry}~\cite{Bardeen:1975zz,1985MNRAS.213..435K}: breaking equatorial-reflection symmetry allows $m_d=\pm1$ components already at $\ell_d=2$, opening the $\Delta m=\pm1$ channel and enabling couplings $\ket{211}\leftrightarrow\ket{210}$ and $\ket{210}\leftrightarrow\ket{21-1}$. In this case all three $m$-sublevels can mix.
\end{itemize}

From an effective-theory viewpoint, it is thus natural to classify disk perturbations by the symmetries they break---axial symmetry controls whether $m_d\neq 0$ channels exist, while equatorial-reflection symmetry controls whether the odd-parity $\ket{210}$ state can participate. Explicit time dependence affects resonance and secular accumulation, but does not modify the selection rule in Eq.~\eqref{eq:md_selection}.

\section{Representative Accretion-Disk Models}
\label{Representative Accretion-Disk Models}

{\color{blue}As discussed in Sec.~\ref{Sec:symmetry}, the symmetry of the disk determines which quadrupolar source multipoles $I_d^{(m_d)}(t)$ are present and hence which $\ket{21m}$ sublevels can mix. We therefore use the symmetry classification derived above as a guide to construct representative disk perturbations. In the ${\ket{211},\ket{210},\ket{21-1}}$ basis, the leading disk-induced Hamiltonian can be schematically written as
\begin{equation}
H(t)=H_0+H_{\mathrm{diag}}(\tilde\Sigma_0)+H_{\mathrm{spiral}}(t;\tilde\Sigma_2)+H_{\mathrm{warp}}(\tilde\Sigma),
\label{eq:Hamiltonian_contributions}
\end{equation}
where $H_{\mathrm{diag}}$ denotes the axisymmetric level shifts, $H_{\mathrm{spiral}}$ represents the equatorially symmetric nonaxisymmetric component that opens the $\ket{211}\leftrightarrow\ket{21-1}$ channel, and $H_{\mathrm{warp}}$ denotes the equatorial-symmetry-breaking component that couples $\ket{210}$ to the even-parity states. Since the disk potential is treated as a weak perturbation, these contributions add linearly at leading order.}

Motivated by this decomposition, we consider two representative disk configurations below. The first is a time-dependent equatorial spiral perturbation, which isolates the $m_d=\pm2$ channel and reduces the dynamics to an effective two-level problem. The second is a small warped disk, which activates the $m_d=\pm1$ channel and produces quasi-static three-level mixing. Their main symmetry properties are summarized in Table~\ref{tab:disk_models}. For the parameter choices or scan ranges used in these examples, the leakage into external states remains perturbative, and the relevant consistency conditions are quantified in Appendix~\ref{app:eta}.

\begin{table}[h]
\caption{\label{tab:disk_models}
Representative disk-perturbation models and their symmetry properties in the quadrupolar ($\ell_d=2$) potential.}
\begin{ruledtabular}
\begin{tabular*}{\linewidth}{@{\extracolsep{\fill}}cccc}
Model & Mass distribution & Geometry & Time dependence \\
\hline
\parbox[t]{0.42\linewidth}{Equatorial Gaussian spiral wave}
& Nonaxisymmetric & Equatorially symmetric & Yes \\
\parbox[t]{0.42\linewidth}{Static warped disk}
& Axisymmetric & Equatorially asymmetric & No \\
\end{tabular*}
\end{ruledtabular}
\end{table}


\subsection{Equatorial Spiral Density Wave with a Gaussian Envelope}
\subsubsection{Equatorial Disk Oscillations}

Guided by the symmetry considerations summarized above, we first consider perturbations that \textit{break axial symmetry} while \textit{preserving equatorial-reflection symmetry}. A geometrically thin equatorial disk can support several oscillation modes when externally perturbed, including eccentric modes~\cite{1983PASJ...35..249K,lubow1991model} and spiral density waves~\cite{1964ApJ...140..646L,Goldreich:1979zz}. The eccentric mode corresponds to azimuthal number $m=1$, while spiral waves may occur for a range of $m$. Since we restrict to the quadrupolar ($\ell_d=2$) gravitational perturbation and the equatorial disk is reflection symmetric, the discussion in Appendix~\ref{C} (around Eq.~\eqref{integration}) implies that odd-$m$ components do not contribute to the relevant matrix elements. We therefore focus on even-$m$ oscillations.

For a general oscillatory surface density, only Fourier components with $|m|\le 2$ can couple to the $\ell_d=2$ potential. It suffices to expand the surface density up to second order,
\begin{equation}
    \Sigma(t,r',\phi')=\sum_{|m|\leq2}\Sigma_m(t,r')\,e^{- \i m\phi'} ,
\end{equation}
where the $m=0$ term describes the axisymmetric background disk and $m\neq 0$ encode nonaxisymmetric perturbations. Reality of $\Sigma$ requires $\Sigma_{-m}=\Sigma_m^\ast$.

The relative importance of the disk's secular evolution can be assessed by comparing the viscous timescale $\tau_\nu\sim R^2/\nu$ with the local Keplerian timescale $\tau_K\sim 1/\Omega_K(R)$~\cite{1981ARA&A..19..137P}. Using $\nu\sim\alpha_\nu c_s H$ and the thin-disk scaling $H/R\sim c_s/v_K=c_s/(R\Omega_K)$~\cite{novikov1973astrophysics}, one finds
\begin{equation}
    \frac{\tau_\nu}{\tau_K}\sim\frac{R^2\Omega_K}{\nu}
    =\frac{1}{\alpha_\nu}\left(\frac{R}{H}\right)^2\gg1 .
\end{equation}
Thus the axisymmetric background may be treated as time independent over the dynamical time of interest,
\begin{equation}
    \Sigma(t,r',\phi')=\Sigma_0(r')+\sum_{0<|m|\leq2}\Sigma_m(t,r')\,e^{- \i m\phi'} .
    \label{Fourier}
\end{equation}

With Eqs.~\eqref{Fourier} and \eqref{eq:H_mm_final}, the leading perturbation Hamiltonian in the $\{\ket{211},\ket{210},\ket{21-1}\}$ basis takes the form
\begin{align}
H_{\mathrm{diag}}(\tilde\Sigma_0)
&=\kappa
\begin{pmatrix}
    -3\tilde\Sigma_0 & 0 & 0 \\
    0 & 6\tilde\Sigma_0 & 0 \\
    0 & 0 & -3\tilde\Sigma_0
\end{pmatrix},\quad
H_{\mathrm{spiral}}(t;\tilde\Sigma_2)
=\kappa
\begin{pmatrix}
    0 & 0 & 9\tilde\Sigma_2(t) \\
    0 & 0 & 0 \\
    9\tilde\Sigma_2^\ast(t) & 0 & 0
\end{pmatrix}.
\end{align}
where $\kappa=2\pi\,\frac{M}{\alpha^3}$ and
\begin{equation}
    \tilde\Sigma_0=\int_{r_{\mathrm{in}}}^{r_{\mathrm{out}}}\frac{\Sigma_0(r')}{{r'}^{2}}\,\mathrm{d} r',
    \qquad
    \tilde\Sigma_m(t)=\int_{r_{\mathrm{in}}}^{r_{\mathrm{out}}}\frac{\Sigma_m(t,r')}{{r'}^{2}}\,\mathrm{d} r' .
    \label{tilde Sigma}
\end{equation}
As anticipated from the symmetry discussion, the odd-parity state $\ket{210}$ decouples from $\ket{211}$ and $\ket{21-1}$ for an equatorially symmetric quadrupole perturbation. The problem therefore reduces to an effective two-level system, which we write as a sum of a time-independent part and a time-dependent interaction,
\begin{align}
\begin{aligned}
        &H = H_0+H_{\mathrm{diag}}(\tilde\Sigma_0)+H_{\mathrm{spiral}},\\
    &H_0+H_{\mathrm{diag}}(\tilde\Sigma_0) =
    \begin{pmatrix}
        E+\epsilon_h-3\kappa\tilde\Sigma_0 & 0\\
        0 & E-\epsilon_h-3\kappa\tilde\Sigma_0
    \end{pmatrix},
    \label{bg matrix}\\
    &H_{\mathrm{spiral}}(t)
=
\begin{pmatrix}
    0 & 9\kappa\tilde\Sigma_2(t)\\
    9\kappa\tilde\Sigma_2^\ast(t) & 0
\end{pmatrix}.
\end{aligned}
\end{align}
The time evolution of the amplitudes follows from the Schr\"odinger equation in the interaction picture,
\begin{equation}
    i\frac{\mathrm{d}}{\mathrm{d}t}
    \begin{pmatrix}
        c_g\\ c_d^{(2)}
    \end{pmatrix}
    =
    \begin{pmatrix}
        0 & 9\kappa\tilde\Sigma_2(t)\,e^{2\i\epsilon_h t} \\
        9\kappa\tilde\Sigma_2^\ast(t)\,e^{-2\i\epsilon_h t} & 0
    \end{pmatrix}
    \begin{pmatrix}
        c_g\\ c_d^{(2)}
    \end{pmatrix}.
    \label{int_scalar_solution}
\end{equation}
which determines the level mixing driven by the nonaxisymmetric $m=\pm2$ disk component.

\subsubsection{Time-Dependent Perturbation and State Evolution}

\begin{figure}[h]
    \centering
    \includegraphics[width=0.8\textwidth]{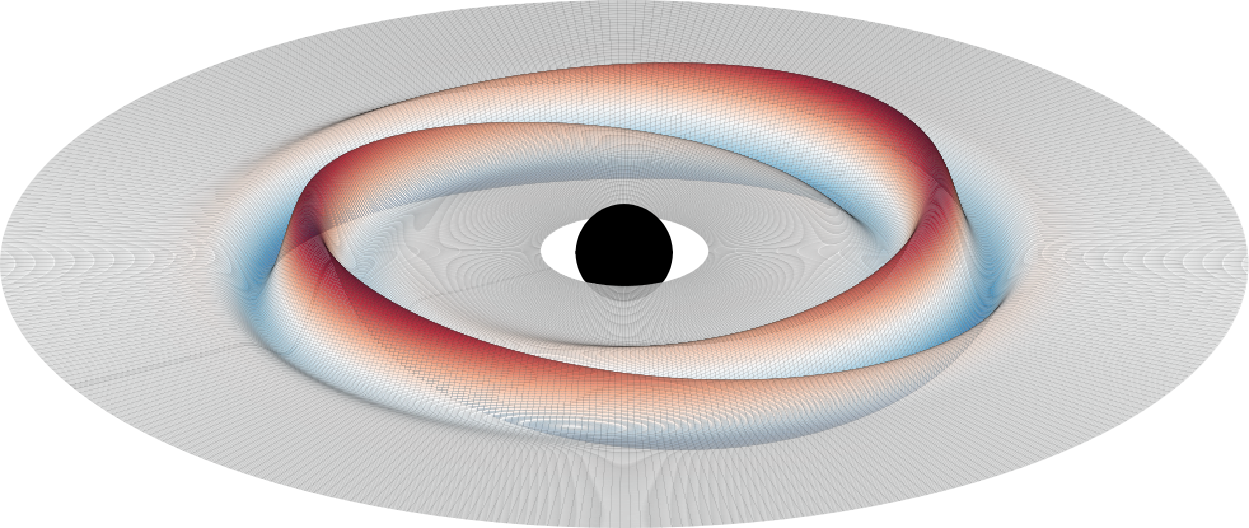}
    \caption{Surface-density profile of an $m=2$ spiral density wave with a Gaussian envelope. The $m=2$ mode produces two spiral arms, while the Gaussian envelope localizes the perturbation radially. The pattern corotates and the wave packet drifts outward.}
\end{figure}

A convenient phenomenological model for the leading nonaxisymmetric
$\ell_d=2$ channel is a localized $m=2$ spiral density-wave packet riding on top
of an axisymmetric background disk. Physically, this describes a two-armed
overdensity pattern whose center drifts radially with a group velocity $v_g$,
so the associated quadrupolar tidal field felt by the cloud turns on when the
packet overlaps the inner disk and then decays as the packet moves outward~\cite{1964ApJ...140..646L,1983PASJ...35..249K,1989ApJ...347..959A}. We
parametrize the surface density as~\cite{2019ApJ...875...37M}
\begin{equation}
    \Sigma(r,\phi,t)
    =
    \Sigma_0(r)
    +
    A\,\frac{r_0}{r}\,
    \exp\!\left[-\frac{(r-r_0-v_g t)^2}{2\sigma_r^2}\right]
    \cos\!\left[2\phi-\bigl(k_r(r-r_0)-\omega_d t\bigr)\right],
    \label{eq:wave}
\end{equation}
where $\Sigma_0(r)$ is the time-independent axisymmetric background, while the
second term is a wave packet centered at $R(t)=r_0+v_g t$ with radial width
$\sigma_r$. The cosine encodes an $m=2$ spiral pattern with radial wavenumber
$k_r$ and pattern frequency $\omega_d$, and $A$ sets the perturbation amplitude. For the benchmark parameters listed in Table~\ref{tab:wave_packet_params}, $r_0$ is chosen near the ISCO so that the perturbation is located in the high-density inner disk, $\sigma_r=\mathcal O(M)$ describes a localized wave packet, and the small value of $v_g$ gives a long coherent driving time. The choices of $k_r$ and $\omega_d$ fix the effective driving frequency seen by the cloud, while $A\sim\Sigma_0$ represents an order-unity density contrast relative to the background disk. These values should therefore be viewed as a favorable coherent-driving benchmark rather than a universal model of turbulent accretion disks.

\begin{table}[t]
\caption{\label{tab:wave_packet_params}
Representative parameters for the Gaussian-envelope $m=2$ spiral-density-wave
packet in Eq.~\eqref{eq:wave}.~\cite{ward1986density,Lin:1986zz,Binney:2008,8201727}}
\begin{ruledtabular}
\begin{tabular*}{\linewidth}{@{\extracolsep{\fill}}ccc}
Parameter & Physical meaning & Representative value \\
\hline
$r_0$ & Initial packet-center radius & $4M$ \\
$v_g$ & Group velocity of the packet & $10^{-8}$ \\
$\sigma_r$ & Characteristic radial width & $M$ \\
$k_r$ & Radial wavenumber (spiral pitch) & $1/M$ \\
$\omega_d$ & Pattern (mode) frequency & $9\times10^{-9}/M$ \\
$A$ & Perturbation amplitude scale & $\sim \Sigma_0$ \\
\end{tabular*}
\end{ruledtabular}
\end{table}

To connect this model to the quadrupolar Hamiltonian, we expand the perturbation
in azimuthal harmonics. Writing the cosine in Eq.~\eqref{eq:wave} as a sum of
exponentials shows that the only nonzero nonaxisymmetric components are $m=\pm2$,
so the relevant Fourier coefficient can be taken as
\begin{equation}
    \Sigma_2(t,r)
    \simeq
    \frac{A}{2}\,\frac{r_0}{r}\,
    \exp\!\left[-\frac{(r-R(t))^2}{2\sigma_r^2}\right]
    \exp\!\left[-\i\bigl(k_r(r-r_0)-\omega_d t\bigr)\right],
    \qquad
    R(t)\equiv r_0+v_g t,
\end{equation}
with $\Sigma_{-2}=\Sigma_2^\ast$ for a real surface density. The disk enters the
two-level problem through the weighted radial moment
\begin{equation}
    \tilde\Sigma_2(t)
    \equiv
    \int_{r_{\mathrm{in}}}^{r_{\mathrm{out}}}
    \frac{\Sigma_2(t,r')}{r'^2}\,\d r'.
    \label{eq:tildeSigma_def_repeat}
\end{equation}
Because the packet is localized around $r'\simeq R(t)$, slowly varying factors
may be evaluated at $R(t)$, while the rapidly varying phase produces a Gaussian
Fourier suppression~\cite{RHEINBOLDT1989544}. This yields the estimate
\begin{equation}
    \tilde\Sigma_2(t)
    \simeq
    \sqrt{\frac{\pi}{2}}\,
    r_0\sigma_r A\,
    e^{-k_r^2\sigma_r^2/2}\,
    \frac{e^{- i\Omega_d t}}{R(t)^3},
    \qquad
    \Omega_d \equiv k_r v_g-\omega_d,
    \label{eq:tildeSigma2}
\end{equation}
where $\Omega_d$ is the \emph{effective driving frequency} seen by the cloud:
the term $k_r v_g$ comes from the radial drift of the packet center, while $\omega_d$ is the intrinsic pattern rotation.
Eq.~\eqref{eq:tildeSigma2} makes the physical control parameters explicit:
the driving strength scales with $A$ and the geometric tidal falloff $R(t)^{-3}$,
the duration is set by $t_{\rm drive}\sim\sigma_r/v_g$, and fine radial structure
is exponentially suppressed by $e^{-k_r^2\sigma_r^2/2}$.

Substituting Eq.~\eqref{eq:tildeSigma2} into the interaction-picture evolution
Eq.~\eqref{int_scalar_solution}, the off-diagonal driving term becomes
\begin{equation}
    9\kappa\,\tilde\Sigma_2(t)\,e^{2i\epsilon_h t}
    \;\simeq\;
    9\kappa\,\tilde A\,\frac{e^{i\delta t}}{R(t)^3},
    \qquad
    R(t)\equiv r_0+v_g t,
    \label{eq:drive_term_delta}
\end{equation}
where we have defined the effective driving detuning
\begin{equation}
    \delta \;\equiv\; 2\epsilon_h-\Omega_d,
    \qquad
    \Omega_d \equiv k_r v_g-\omega_d .
    \label{eq:delta_def}
\end{equation}
Thus the driven two-level dynamics is controlled by a competition between the
maximum mixing strength (set by the peak quadrupolar coupling near $t\simeq0$)
and the detuning $\delta$ (set by the intrinsic level splitting $2\epsilon_h$
relative to the disk’s pattern frequency $\Omega_d$). It is therefore natural
to introduce the dimensionless control parameter~\cite{Binney:2008,1992rcd..book.....L}
\begin{equation}
    \Lambda
    \;\equiv\;
    \frac{\Omega_{\rm mix}(t\simeq0)}{|\delta|}
    \;\simeq\;
    \frac{9\kappa|\tilde A|/r_0^{3}}{\left|2\epsilon_h-(k_r v_g-\omega_d)\right|},
    \label{eq:Lambda_def}
\end{equation}
where $\Omega_{\rm mix}(t)\equiv 9\kappa|\tilde A|/R(t)^3$ is the instantaneous
mixing rate induced by the $m=\pm2$ disk component. In addition, because the
envelope moves outward with speed $v_g$, the driving acts effectively only over
a duration $ t_{\rm drive}$, so smaller $v_g$ corresponds to longer coherent forcing and hence larger
cumulative mixing, even when the instantaneous perturbation is weak.

The impact on the scalar cloud is conveniently quantified by the late-time
effective growth rate
\begin{equation}
    \Gamma_{\rm eff}(t)=|c_g(t)|^2\Gamma_{211}+|c_d(t)|^2\Gamma_{21-1},
\end{equation}
and by the growth-factor ratio
\begin{equation}
    \chi \equiv \frac{\Gamma_{\rm eff}(\infty)}{\Gamma_{\rm eff}(0)} .
\end{equation}
For the superradiant initial condition $c_g(0)=1$, $c_d(0)=0$ one has
$\Gamma_{\rm eff}(0)=\Gamma_{211}$, and the late-time ratio can be written
directly in terms of the transferred population
$P_d\equiv |c_d(\infty)|^2$,
\begin{equation}
    \chi
    =
    \frac{\Gamma_{211}(1-P_d)+\Gamma_{21-1}P_d}{\Gamma_{211}}
    =
    1-\left(1-\frac{\Gamma_{21-1}}{\Gamma_{211}}\right)P_d .
    \label{eq:chi_Pd}
\end{equation}
Eq.~\eqref{eq:chi_Pd} makes the physics transparent: disk driving does not
modify the microscopic rates $\Gamma_{211}$ and $\Gamma_{21-1}$, but it can
redistribute probability between the growing and decaying levels; once $P_d$
becomes ${\cal O}(1)$, the decaying component can dominate and quench
superradiance ($\chi<0$).

In practice, the mixing strength $P_d$ is controlled by the single ratio
$\Lambda$ (enhanced by a long driving time $t_{\rm drive}$). This motivates the
following interpretation of the $(\alpha,\Sigma_0)$ scan in Fig.~\ref{Gamma}:
\begin{itemize}
    \item $\Lambda\ll 1$: weak, off-resonant driving $\Rightarrow P_d\ll1$ and
    $\chi\simeq1$ (negligible mixing).
    \item $\Lambda\sim 1$: comparable coupling and detuning $\Rightarrow
    P_d={\cal O}(0.1\text{--}1)$ and $0<\chi<1$ (suppressed growth).
    \item $\Lambda\gg 1$ (especially when $|\delta|\lesssim t_{\rm drive}^{-1}$):
    strong/near-resonant transfer $\Rightarrow P_d\to{\cal O}(1)$ and $\chi<0$
    (quench region).
\end{itemize}
The parametric dependence is also clear from Eq.~\eqref{eq:Lambda_def}:
$\alpha$ enters mainly through the intrinsic splitting $\epsilon_h(\alpha)$ and
the prefactor $\kappa\propto M/\alpha^3$, while the disk strength enters through
$|\tilde A|\propto A$ and the geometric tidal falloff $r_0^{-3}$; finally, the
duration of coherent forcing scales as $t_{\rm drive}$, making
small $v_g$ particularly efficient at accumulating mixing.

\begin{figure}[h]
    \centering
    \includegraphics[width=0.75\linewidth]{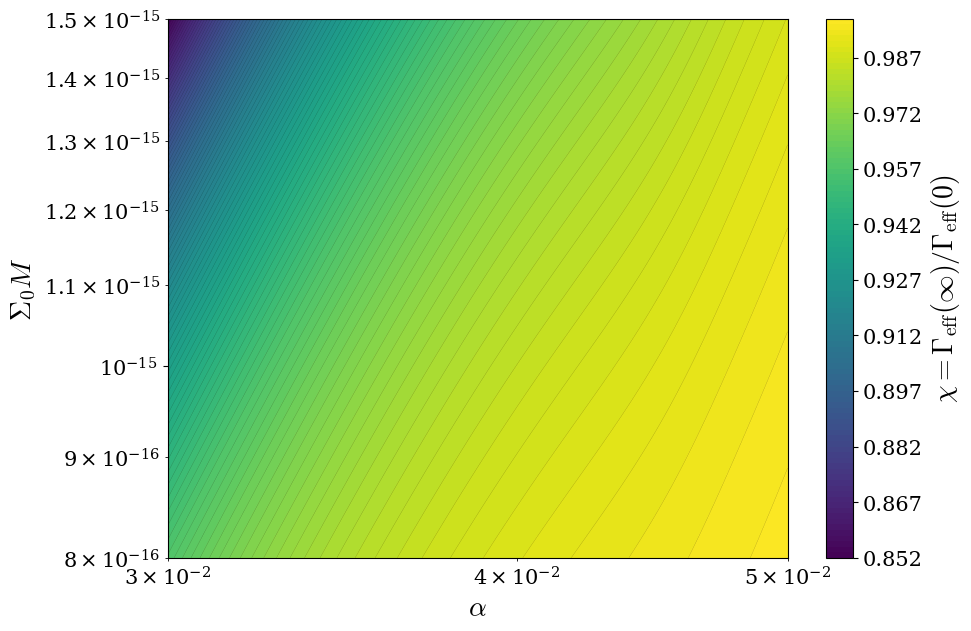}
    \caption{The figure shows the scanned values of the growth-factor ratio $\chi \equiv \Gamma_{\mathrm{eff}}(\infty)/\Gamma_{\mathrm{eff}}(0)$in the $(\alpha,\Sigma_0)$ plane. A larger surface density and a smaller $\alpha$ correspond to a stronger suppression of the superradiant effect by the accretion disk.}
    \label{Gamma}
\end{figure}

The results are summarized in Fig.~\ref{Gamma}. For the stellar-mass thin-disk parameters considered here, the region mainly falls into two regimes. When the driving is weak or sufficiently off resonance, the transferred population remains small and $\chi\simeq1$. In the stronger-mixing region, the spiral perturbation transfers part of the cloud population from the growing state $\ket{211}$ to the decaying state $\ket{21-1}$, leading to suppressed growth with $0<\chi<1$. 
\color{blue}
The absence of a broad quenching region with $\chi<0$ reflects the small self-gravity
of typical stellar-mass thin disks: the surface-density normalization
$\Sigma_0M\simeq10^{-17}$ strongly suppresses $\Omega_{\rm mix}$, so the spiral
perturbation can reduce $\Gamma_{\rm eff}$ but is generally not strong enough for
the decaying component to dominate the late-time state. In addition, for the stellar-mass thin-disk
benchmark used in Fig.~\ref{Gamma}, the near-resonant region overlaps the multipole-validity
domain only in the narrow range$\alpha\simeq 0.03\text{-}0.05$.
For larger $\alpha$, in particular for the phenomenologically relevant region
$\alpha\gtrsim 0.1$, the system is sufficiently off resonance and the mixing becomes
negligible.

\color{black}

\color{blue}
The above result also depends on the phase coherence of the nonaxisymmetric $m=2$ component. The Gaussian-envelope packet in Eq.~\eqref{eq:wave} represents the coherent limit, in which the phase of $\widetilde\Sigma_2(t)$ remains correlated over the effective driving time $t_{\rm drive}$. To estimate the effect of finite coherence, one may write the transition amplitude as
\begin{equation}
c_d(\infty)
\sim
-\i\int_0^{t_{\rm drive}} \d t
\Omega_{\rm mix}(t)
e^{\i(\delta t+\phi(t))} ,
\end{equation}
where $\phi(t)$ denotes stochastic phase fluctuations. Near resonance, coherent accumulation gives
\begin{equation}
P_d^{\rm coh}\sim \Omega_{\rm mix}^2t_{\rm drive}^2 .
\end{equation}
If the phase decorrelates after a shorter time $\tau_{\rm coh}\ll t_{\rm drive}$, the driving can be viewed as $N\sim t_{\rm drive}/\tau_{\rm coh}$ independent coherent segments. Each segment contributes $\Delta c_d\sim\Omega_{\rm mix}\tau_{\rm coh}$, so
\begin{equation}
P_d
\sim
N|\Delta c_d|^2
\sim
\Omega_{\rm mix}^2t_{\rm drive}\tau_{\rm coh}
=
P_d^{\rm coh}\frac{\tau_{\rm coh}}{t_{\rm drive}},
\end{equation}
up to detuning-dependent factors. Finite coherence therefore reduces the cumulative population transfer, but does not change the underlying $\Delta m=\pm2$ selection channel.
\color{black}

In summary, the equatorial spiral perturbation provides a time-dependent channel for transferring population from the growing mode to the decaying mode. For the stellar-mass thin disks considered here, this mainly leads to partial suppression of superradiance, while complete quenching requires stronger and sufficiently coherent disk perturbations. The quenching region should therefore be interpreted as a favorable coherent upper envelope rather than a generic prediction for turbulent disks.

}

\subsection{Static Warped Disk: Equatorial-Symmetry Breaking}

\subsubsection{Geometry and the $\Delta m=\pm1$ Coupling Channel}

We next consider perturbations that preserve axial symmetry but break equatorial-reflection symmetry, so that the $m_d=\pm1$ channel is activated and the odd-parity state $\ket{210}$ can mix with $\ket{211}$ and $\ket{21-1}$. A tilted disk is generic if the angular-momentum axis of the inflowing material is misaligned with the BH spin. In a Kerr spacetime, frame dragging tends to align the inner disk with the BH equatorial plane while the outer disk remains tilted, producing a warped transition region (the Bardeen--Petterson effect)~\cite{Bardeen:1975zz,10.1093/mnras/202.4.1181}. The local tilt is described by an inclination angle $\beta(r)$. In addition, Lense--Thirring precession causes rings at different radius to precess at different rates. The resulting differential precession exerts a twisting torque which, if too large, can even tear the disk~\cite{dougan2018instability,nixon2012tearing}. When viscous stresses are able to counteract this torque, the twist saturates~\cite{ogilvie1999non}; we parametrize the residual twist by a phase $\gamma(r)$. For our purposes, only the leading effect of the warp on the quadrupolar potential is retained.

\begin{figure}[h]
\centering
\includegraphics[width=0.5\textwidth]{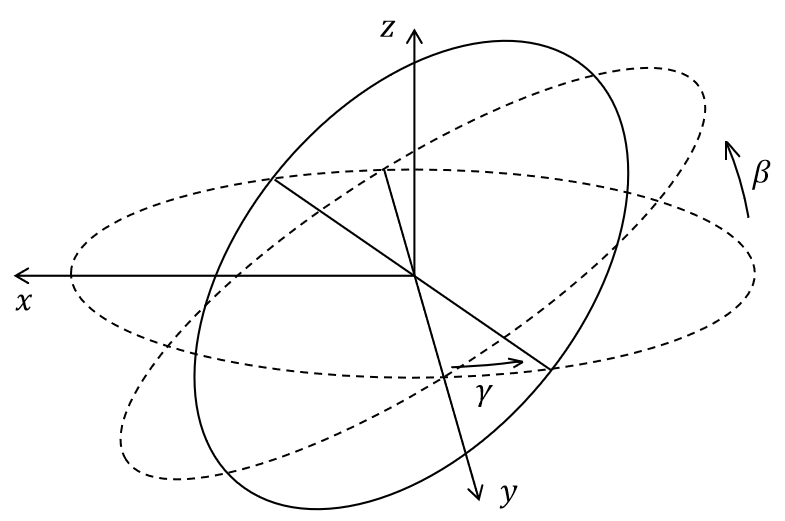}\hfill
\includegraphics[width=0.35\linewidth]{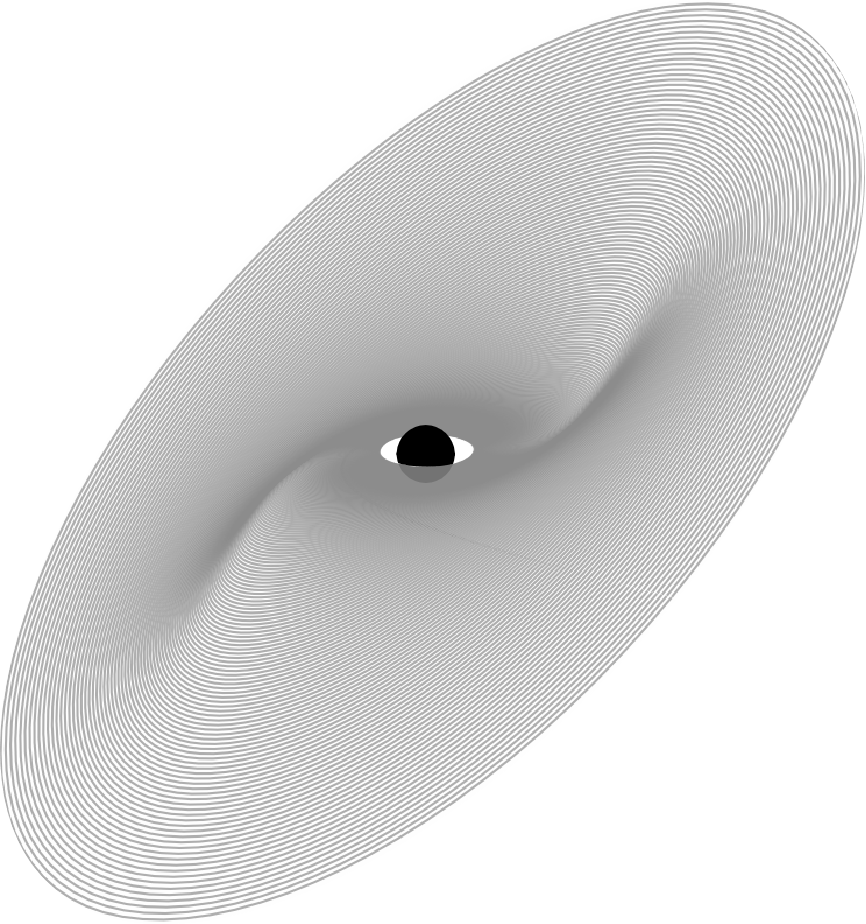}
\caption{Accretion-disk geometry and notation. Left: Geometric construction of a thin warped disk. Each ring at radius $r$ is obtained by tilting the corresponding equatorial ring by $\beta(r)$ and applying a twist $\gamma(r)$; the small-tilt limit $\beta\ll1$ is assumed~\cite{10.1093/mnras/202.4.1181,10.1093/mnras/258.4.811}. Right: schematic warped configuration in which the inner disk tends to align with the BH equatorial plane while the outer disk remains tilted, producing a warped transition region.}
\label{fig:disk_geometry}
\end{figure}

\subsubsection{Quasi-Static Warp Approximation}

An important simplification is that the warped configuration can be treated as
\emph{effectively static} on the timescales relevant for the cloud evolution.
Physically, any coherent Lense--Thirring precession of a global warp is expected
to be strongly damped by viscous dissipation, so that the disk relaxes to a
quasi-steady geometry rather than maintaining a long-lived precessing mode.
This statement can be quantified in the Papaloizou--Pringle rigid-disk
approximation~\cite{10.1093/mnras/202.4.1181}, where viscous communication across the disk is fast enough that
the warp responds coherently and the damping time $\tau_{\mathrm{damp}}$ can be
compared directly to the global precession period $T_p$~\cite{lodato2007warp}.  {\color{blue}
Using Eq.~\eqref{precession} in Appendix~\ref{D}, one finds
\begin{equation}
     \frac{\tau_{\mathrm{damp}}}{T_p}
     \sim
     \frac{2\tilde a}{\alpha_2 h^2}
     F_p(x)
     \left(\frac{M}{r_{\mathrm{in}}}\right)^{3/2}
     x^{3/4}.
\end{equation}
For representative parameters $\alpha=0.1$, $\tilde a=0.9$, $r_{\mathrm{in}}\sim 2.32M$,
$r_{\mathrm{out}}\sim 10^5M$, and $h\sim10^{-2}$, together with
$\alpha_2\simeq 1/(2\alpha)$ in this approximation, this estimate gives
\begin{equation}
    \frac{\tau_{\mathrm{damp}}}{T_p}\approx 3.3\times10^{-2} \ll 1 ,
\end{equation}
implying that any global precession decays within less than a cycle and the disk
settles into a steady warp.} In the language of the source multipoles $I_d^{(m_d)}(t)$, this conclusion can be
phrased more directly: a warped configuration corresponds to an ultra--low--frequency
contribution, i.e.\ the $\omega\to0$ limit of a slowly varying perturbation.
Therefore, on the timescales relevant for the cloud's linear-growth stage, the
warp acts as a quasi-static background. One may schematically parametrize any
residual relaxation as
\begin{equation}
    \tilde\Sigma(t)\simeq \tilde\Sigma\e^{-t/\tau_{\mathrm{damp}}},
\end{equation}
but for the present purposes it is sufficient to treat $\tilde\Sigma$ as constant
over the timescale of interest.

\subsubsection{Static Gravitational Perturbation}

To isolate the warp-induced effect, we assume an axisymmetric and time-independent surface density,
so that only the $m_d=0$ (diagonal) and $m_d=\pm1,\pm2$ (warp) source multipoles are present at $\ell_d=2$.
Following the decomposition in Eq.~\eqref{eq:Hamiltonian_contributions}, we write the Hamiltonian in the
$\{\ket{211},\ket{210},\ket{21-1}\}$ basis as
\begin{equation}
    H \;=\; H_0 \;+\; H_{\mathrm{diag}}(\tilde\Sigma_0)\;+\; H_{\mathrm{warp}}(\tilde\Sigma),
\end{equation}
where the axisymmetric surface-density moment is
\begin{equation}
    \tilde\Sigma_0=\int_{r_{\mathrm{in}}}^{r_{\mathrm{out}}}\frac{\Sigma(r')}{{r'}^2}\,\mathrm{d}r',
\end{equation}
and the warp geometry is encoded in the weighted moment
\begin{equation}
    \tilde\Sigma=\int_{r_{\mathrm{in}}}^{r_{\mathrm{out}}}\frac{\Sigma(r')}{{r'}^2}\,
    \beta(r')\,e^{-i\gamma(r')}\,\mathrm{d}r' .
    \label{eq:weighted moment}
\end{equation}
Keeping only terms linear in the small tilt $\beta\ll1$ (see Appendix~\ref{E} for details),
the diagonal contribution (from the axisymmetric component $m_d=0$) takes the form
\begin{equation}
    H_0 + H_{\mathrm{diag}}(\tilde\Sigma_0)=
    \begin{pmatrix}
          E+\epsilon_{h}-3\kappa\tilde\Sigma_0 & 0 & 0 \\
        0 &   E+6\kappa\tilde\Sigma_0 & 0 \\
        0 & 0 &   E-\epsilon_{h}-3\kappa\tilde\Sigma_0
    \end{pmatrix},
\end{equation}
while the warp-induced (equatorially asymmetric) part is
\begin{equation}
    H_{\mathrm{warp}}(\tilde\Sigma)=\frac{9\sqrt{2}}{2}\kappa
    \begin{pmatrix}
        0 & -\tilde\Sigma & 0 \\
        -\tilde\Sigma^\ast & 0 & \tilde\Sigma\\
        0 & \tilde\Sigma^\ast & 0
    \end{pmatrix}.
\end{equation}
Notably, for a thin ring with a small local tilt $\beta(r')\ll 1$ (Appendix~\ref{E}), one finds the warp amplitude in the linear order,
\begin{equation}
    I_d^{(0)}={\cal O}(1),
    \qquad
    I_d^{(\pm1)}={\cal O}(\beta)\,e^{\mp \i\gamma},
    \qquad
    I_d^{(\pm2)}={\cal O}(\beta^2),
\end{equation}
so a small warp activates the $m_d=\pm1$ channel already at $\ell_d=2$, whereas
the $m_d=\pm2$ channel is subleading in the linear-tilt approximation.

Because the warped-disk perturbation is (quasi-)time independent, it does not
provide a coherent \emph{drive} as in the spiral case; instead it acts as a
\emph{static basis rotation}. The Hamiltonian is constant, so the cloud
reorganizes into new stationary eigen-combinations that are mixtures of the
unperturbed $\{\ket{211},\ket{210},\ket{21-1}\}$ sublevels. The microscopic
growth/decay rates $\Gamma_{21m}$ are unchanged; any change in the effective
growth rate therefore comes only from a reshuffling of probability among the
three sublevels. In stationary perturbation theory this reshuffling is
controlled by the ratio of the off-diagonal warp coupling
$\bar\Sigma\equiv\frac{9\sqrt2}{2}\kappa\tilde\Sigma$ to the relevant diagonal
splittings of $H_0+H_{\rm diag}$, namely the $m=+1\leftrightarrow 0$ and
$0\leftrightarrow -1$ gaps
\begin{align}
\begin{aligned}
\Delta_{g0}\equiv (E+\epsilon_h-3\kappa\tilde\Sigma_0)-(E+6\kappa\tilde\Sigma_0)
=\epsilon_h-9\kappa\tilde\Sigma_0,\\
\Delta_{0d}\equiv (E+6\kappa\tilde\Sigma_0)-(E-\epsilon_h-3\kappa\tilde\Sigma_0)
=\epsilon_h+9\kappa\tilde\Sigma_0 .
\end{aligned}
\end{align}
To leading order, the perturbed cloud state can be written as
\begin{equation}
\ket{\psi_c}=
\left(c_g+\frac{\bar{\Sigma}}{\Delta_{g0}}c_d^{(1)}\right)\ket{211}
+\left(c_d^{(1)}-\frac{\bar{\Sigma}^\ast}{\Delta_{g0}}c_g-\frac{\bar{\Sigma}}{\Delta_{0d}}c_d^{(2)}\right)\ket{210}
+\left(c_d^{(2)}+\frac{\bar{\Sigma}^\ast}{\Delta_{0d}}c_d^{(1)}\right)\ket{21-1},
\label{eq:perturbed_state}
\end{equation}
so the mixing is
enhanced when one of the splittings is accidentally suppressed. In particular,
it is convenient to define the resonance indicator
\begin{align}
    \Delta_{\rm res}(\alpha)\equiv \Delta_{g0}=\epsilon_h(\alpha)-9\kappa(\alpha)\tilde\Sigma_0,
\end{align}
which measures how close the $\ket{211}$ and $\ket{210}$ diagonal levels are
after including the axisymmetric disk field; 
{\color{blue}when $\Delta_{\rm res}\simeq 0$, namely
\begin{equation}
    \alpha_{\rm res}=\left[\frac{216\pi}{\tilde a}(\tilde\Sigma_0M^2)\right]^{\frac{1}{9}}
\end{equation}
the $\ket{211}\leftrightarrow\ket{210}$ admixture in
Eq.~\eqref{eq:perturbed_state} is strongly enhanced and any initial state with
access to $\ket{210}$ can be substantially reshuffled.

We adopt a representative model for the warped accretion disk,
\begin{align}
\Sigma(r)
&=
\frac{10^{-15}}{M}
\left(\frac{r}{M}\right)^{-3/4},
\\
\beta(r)
&=
0.2\left(1+\frac{r}{R_{\rm BP}}\right)^{-1},
\\
\gamma(r)
&=
0.5\ln\left(\frac{r}{r_{\rm in}}\right),
\\
R_{\rm BP}
&=
M
\left(\frac{\tilde a}{\alpha_\nu}\right)^{2/3}
\left(\frac{H}{R}\right)^{-4/3}.
\end{align}
Together with $\alpha_\nu=0.1$ and $H/R=0.01$, this model is used to produce
Fig.~\ref{Fig:Gamma_eff}, where we compare the effective growth rates for several representative initial-state compositions.
\begin{figure}[htbp]
    \centering
    \begin{tabular}{c}
        \includegraphics[width=0.95\linewidth]{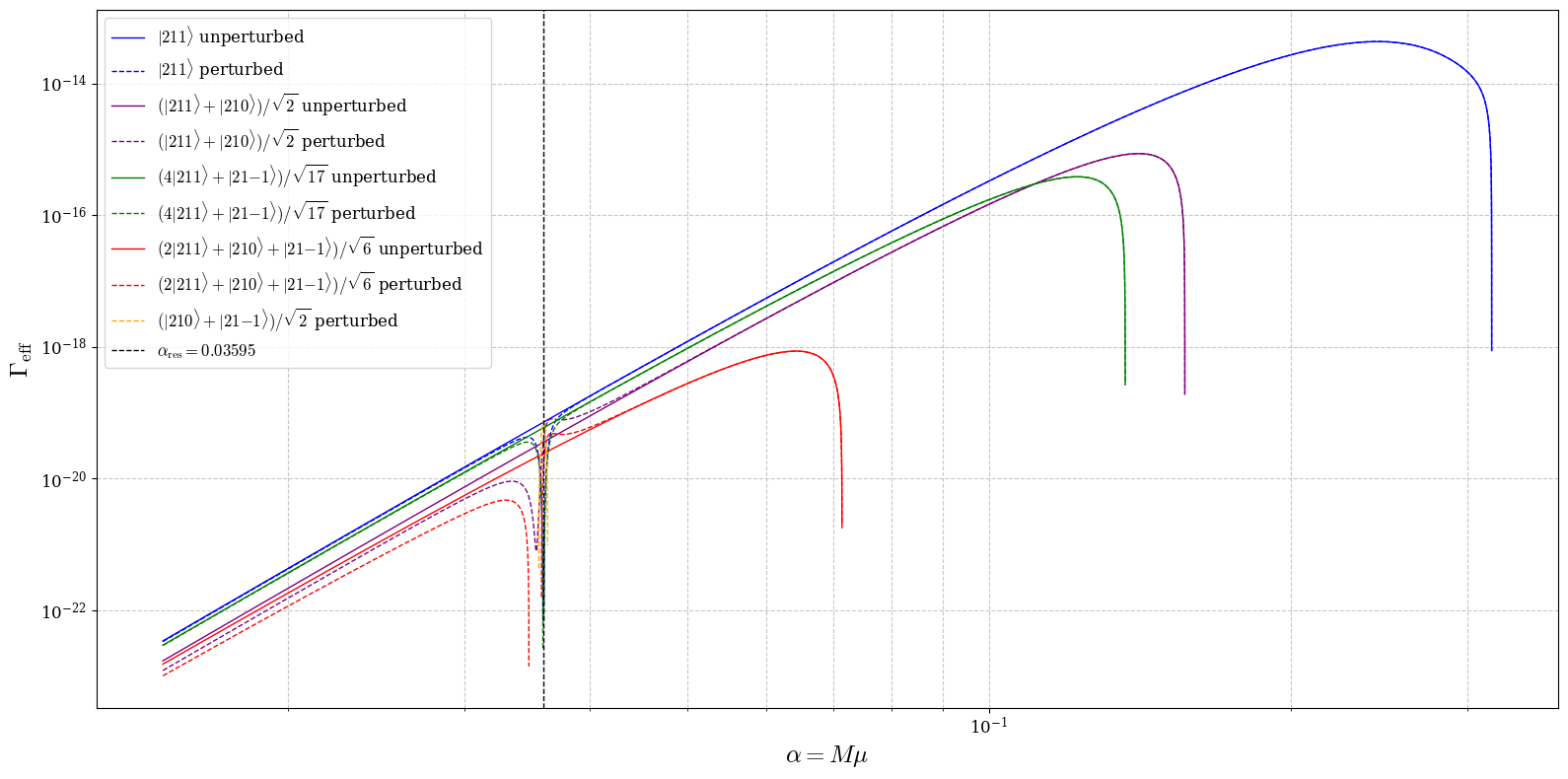} \\
        (a) Full scan range \\
        \includegraphics[width=0.95\linewidth]{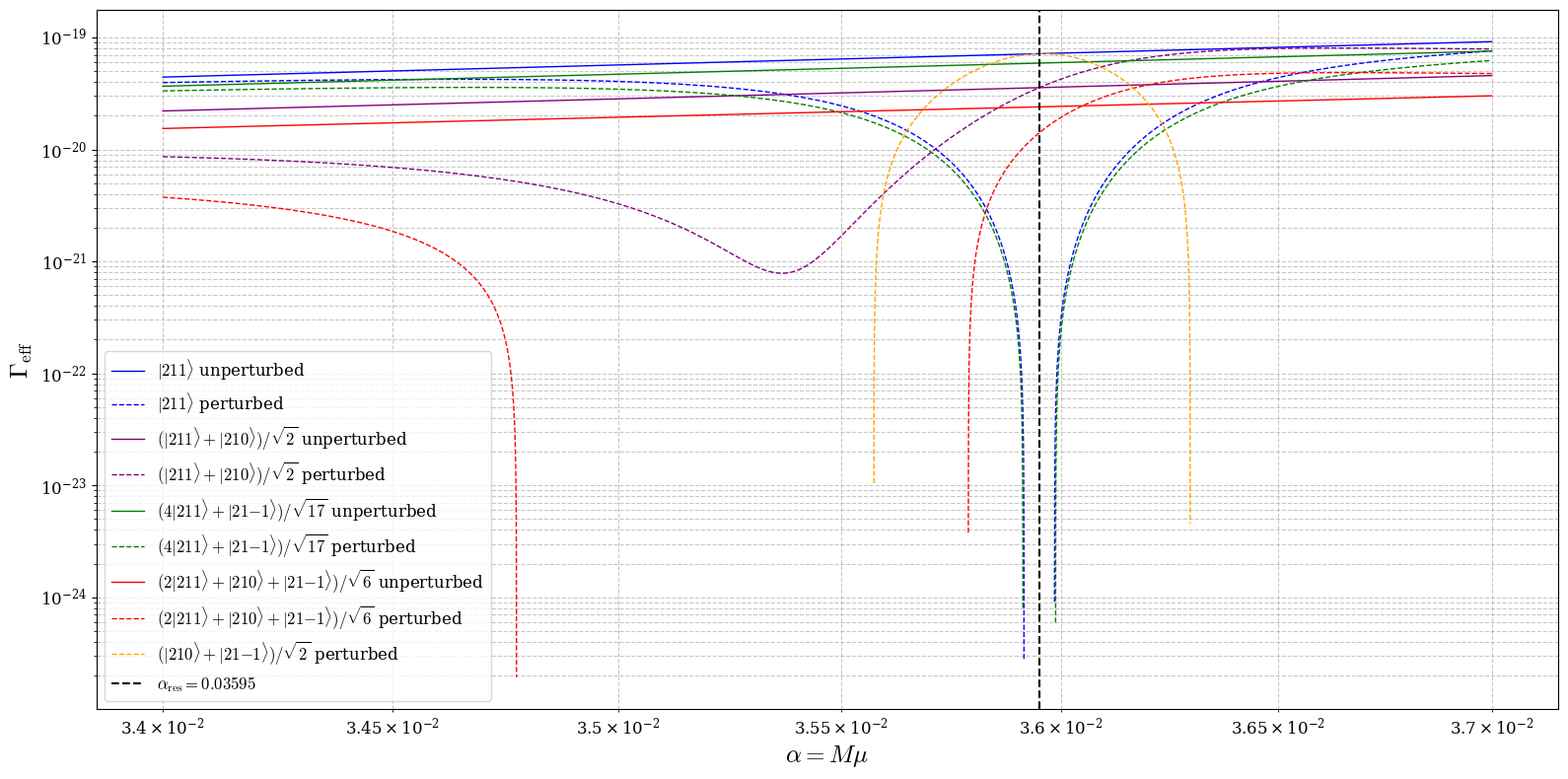} \\
        (b) Region where the non-degenerate condition breaks down
    \end{tabular}
    \caption{Effective growth rate $\Gamma_{\mathrm{eff}}$ as a function of the
    gravitational fine-structure constant $\alpha$ in the presence of a static
    warped-disk perturbation. (a) shows the full scan range, illustrating
    the overall dependence of $\Gamma_{\mathrm{eff}}$ on $\alpha$ for different
    initial-state compositions. (b) zooms into the region where the
    non-degenerate perturbative condition breaks down, namely where the
    diagonal splitting between $\ket{211}$ and $\ket{210}$ is strongly
    suppressed and static mixing becomes most efficient. Colors label the
    unperturbed initial-state compositions
    $(c_g,c_d^{(1)},c_d^{(2)})=(1,0,0)$ (blue),
    $(1/\sqrt2,\,1/\sqrt2,\,0)$ (purple),
    $(4/\sqrt{17},\,0,\,1/\sqrt{17})$ (green), 
    $(2/\sqrt6,\,1/\sqrt6,\,1/\sqrt6)$ (red), and $(0,1/\sqrt{2},1/\sqrt{2})$(orange). Solid curves show the unperturbed
    results, while dashed curves include the warp-induced mixing. The vertical
    dashed line marks $\Delta_{\rm res}(\alpha)=0$, with
    $\Delta_{\rm res}\equiv \epsilon_h-9\kappa\tilde\Sigma_0$, where the
    diagonal splitting between $\ket{211}$ and $\ket{210}$ is suppressed and the
    static mixing is maximized.}
    \label{Fig:Gamma_eff}
\end{figure}

Since the perturbation induced by a stellar-mass thin disk is weak, the warp effect is mainly visible near the accidental near-degeneracy shown in Fig.~\ref{Fig:Gamma_eff}(b). Around $\alpha\simeq\alpha_{\rm res}$, the diagonal separation
$\Delta_{\rm res}=\epsilon_h-9\kappa\widetilde\Sigma_0$ between $\ket{211}$ and $\ket{210}$ is strongly suppressed, so the non-degenerate perturbative condition breaks down. Quantitatively reliable results in this narrow region require diagonalizing the near-degenerate subspace, or equivalently solving the full projected three-level Hamiltonian. Nevertheless, the qualitative effect is clear: the static warp reshuffles the cloud among growing and decaying components. For an initially growing state, part of the weight is rotated into the decaying sector and $\Gamma_{\rm eff}$ is suppressed. For initially decaying or mixed states, the same mixing can rotate part of the state into growing-dominated eigen-combinations and enhance $\Gamma_{\rm eff}$.

This state dependence also explains the mild enhancement of some growing-state curves slightly above $\alpha_{\rm res}$ and the displacement of the suppression windows relative to $\Delta_{\rm res}=0$. The warp does not induce a one-way transfer from the growing state to the decaying state; rather, it changes the perturbed eigenbasis. To see how the shift arises, consider the perturbed effective growth rate
\begin{equation}
\Gamma_{\rm eff}^{\rm pert}
=
\frac{
|A_g|^2\Gamma_{211}
+
|A_0|^2\Gamma_{210}
+
|A_d|^2\Gamma_{21-1}
}{
|A_g|^2+|A_0|^2+|A_d|^2
},
\end{equation}
where $A_g,A_0,A_d$ follow from Eq.~\eqref{eq:perturbed_state}. Keeping only the dominant $\ket{211}\leftrightarrow\ket{210}$ channel, with $\delta\equiv\Delta_{g0}$ and $c_d=0$, the numerator contains
\begin{equation}
\mathcal N
=
\Gamma_g |c_g|^2
+
\Gamma_0 |c_0|^2
+
\frac{|\bar\Sigma|^2}{\delta^2}
\left(\Gamma_g |c_0|^2+\Gamma_0 |c_g|^2\right)
+
2(\Gamma_g-\Gamma_0)
\mathrm{Re}\left(
\frac{\bar\Sigma}{\delta}c_0 c_g^\ast
\right).
\end{equation}
The last term is odd under $\delta\to-\delta$. Hence, when the initial state contains both $\ket{211}$ and $\ket{210}$ components, this interference term shifts the minimum of $\Gamma_{\rm eff}$ away from $\delta=0$, as seen for the red and purple curves. If $c_0=0$, the interference term vanishes and the suppression window remains aligned with $\alpha_{\rm res}$, as in the blue and green curves.

In summary, for the stellar-mass thin-disk parameters considered here, a quasi-static warp affects the effective growth rate mainly in a narrow near-resonant region. It can suppress superradiance for initially growing states, but it can also enhance $\Gamma_{\rm eff}$ for decaying or mixed initial states. The sign and size of the effect are therefore controlled not only by the mixing strength, but also by the projection of the initial cloud state onto the perturbed eigenbasis.

The above discussion assumes a quasi-static warp. However, when the aforementioned assumption of a quasi-static warp breaks down, one may also consider a periodically precessing warp (which may be sustained by external torques, disk self-gravity, or global disk modes) as a complementary coherent-driving limit. In this scenario, the periodic precession primarily shifts the rotating-frame diagonal energies and replaces the static near-degeneracy conditions with frequency-matching resonance conditions. Crucially, because these resonance conditions cannot generally be satisfied simultaneously for both adjacent channels, a precessing warp can selectively enhance one specific mixing channel without activating the other. We relegate a detailed derivation of this coherent-driving limit and the corresponding rotating-frame transformation to Appendix~\ref{app:precessing_warp}.

}

\section{Conclusion and discussion}
\label{sec:conclusion}

We have studied how accretion-disk self-gravity modifies Kerr BH superradiance and the evolution of gravitational atoms. For stellar-mass BHs surrounded by geometrically thin disks, the disk can be treated as a weak external perturbation during the linear superradiant growth stage. In the nearly degenerate $n=2$ subspace ${\ket{211},\ket{210},\ket{21{-}1}}$, the disk perturbation does not directly change the microscopic rates $\Gamma_{21m}$. Instead, it redistributes the cloud population among growing and decaying modes, thereby changing the effective growth rate $\Gamma_{\rm eff}$.

The level mixing is mainly controlled by symmetry. In the freely falling frame, the monopole only shifts the energy origin and the dipole vanishes, so the quadrupole $\ell_d=2$ gives the leading nontrivial tidal perturbation. The disk geometry enters through the source multipoles $I_d^{(m_d)}$, while the selection rule $m_d=m'-m$ determines which magnetic sublevels can mix. Axisymmetric disks only shift the diagonal energies. Equatorially symmetric but nonaxisymmetric perturbations open the $m_d=\pm2$ channel and mix $\ket{211}$ with $\ket{21{-}1}$. Breaking equatorial-reflection symmetry opens the $m_d=\pm1$ channel and allows $\ket{210}$ to participate, leading to genuine three-level mixing.

{\color{blue}
The first representative model is an equatorial Gaussian-envelope $m=2$ spiral density wave. Since this perturbation preserves equatorial reflection symmetry, the odd-parity state $\ket{210}$ decouples and the system reduces to a driven two-level problem involving $\ket{211}$ and $\ket{21{-}1}$. The relevant competition is between the mixing rate and the effective detuning $\delta=2\epsilon_h-\Omega_d$, or equivalently the ratio $\Lambda\sim\Omega_{\rm mix}/|\delta|$. For the representative stellar-mass thin-disk parameters considered here, the main outcome is partial suppression, with $0<\chi<1$, where $\chi=\Gamma_{\rm eff}(\infty)/\Gamma_{\rm eff}(0)$. Physically, the spiral wave transfers part of the population from the growing mode into the decaying mode, reducing the effective superradiant growth rate. A complete quench with $\chi<0$ would require stronger disk self-gravity, better frequency matching, or denser disk environments.

This spiral-wave result should be regarded as a coherent-driving benchmark. If the $m=2$ perturbation remains phase coherent over the driving time, the transferred population scales as $P_d^{\rm coh}\sim\Omega_{\rm mix}^2t_{\rm drive}^2$. If the perturbation decorrelates on a shorter time $\tau_{\rm coh}\ll t_{\rm drive}$, the accumulation becomes random-walk-like, $P_d\sim\Omega_{\rm mix}^2t_{\rm drive}\tau_{\rm coh}$. Thus finite coherence does not change the $m_d=\pm2$ selection channel, but it weakens the accumulated population transfer and moves the system toward the weak-mixing limit $\chi\simeq1$.

The second representative model is a warped disk. A small local tilt breaks equatorial-reflection symmetry and activates the $m_d=\pm1$ channel at linear order in the tilt angle. In the quasi-static limit, the warp does not act as a resonant drive. Instead, it rotates the stationary eigen-combinations within the three-level subspace, so the change in $\Gamma_{\rm eff}$ is controlled by the projection of the initial cloud state onto the perturbed eigenbasis.

For the stellar-mass thin-disk parameters adopted here, the warp effect is concentrated near the accidental near-degeneracy
\begin{equation}
\Delta_{\rm res}(\alpha)
\equiv
\epsilon_h(\alpha)-9\kappa(\alpha)\widetilde{\Sigma}_0
\simeq 0 .
\end{equation}
Near this point, the diagonal separation between $\ket{211}$ and $\ket{210}$ is strongly suppressed, so even a weak warp can produce efficient mixing. For initially growing states, the warp transfers part of the weight into decaying components and suppresses superradiant growth. For initially decaying or mixed states, the same static reshuffling can rotate weight into growing-dominated eigen-combinations and enhance $\Gamma_{\rm eff}$. Therefore, a warped disk does not simply suppress superradiance in a one-way manner; its effect is state dependent.

This state dependence also explains why the strongest suppression need not occur exactly at $\Delta_{\rm res}=0$. If the initial state contains coherent components of both $\ket{211}$ and $\ket{210}$, interference between the original amplitude and the warp-induced admixture can shift the minimum of $\Gamma_{\rm eff}$ away from the nominal near-degeneracy point. Thus the location and depth of the growth gap are determined not only by the diagonal level crossing, but also by the relative amplitudes and phases in the initial cloud state.

As a complementary limit, we also considered coherent periodic precession of the warp. A rotating-frame transformation maps the time-dependent problem to an effective static Hamiltonian with shifted diagonal energies. The static near-degeneracy condition is then replaced by
\begin{align}
\ket{211}\leftrightarrow\ket{210}:
\qquad
\Delta_{g0}\simeq s\Omega_p,
\\
\ket{210}\leftrightarrow\ket{21{-}1}:
\qquad
\Delta_{0d}\simeq s\Omega_p .
\end{align}
Thus precession does not change the $m_d=\pm1$ selection rule opened by the warp; it only shifts the resonance condition through the precession frequency and direction. For fixed $s\Omega_p$, the two adjacent channels are not generally resonant at the same time.
}

In summary, accretion-disk gravity can reshape BH superradiance by mixing gravitational-atom levels. A coherent equatorial spiral wave provides a time-dependent channel that transfers population from the growing mode to a decaying mode, leading mainly to suppression for stellar-mass thin disks and possible quenching only in stronger and sufficiently coherent environments. A quasi-static warp produces state-dependent three-level mixing, generating narrow growth gaps near accidental near-degeneracies and possible local enhancement for suitable initial states. These effects are most important near resonant or near-degenerate regions, where the small intrinsic level splittings of the gravitational atom amplify otherwise weak environmental perturbations.

Future work should include larger level subspaces, higher disk multipoles, more realistic disk dynamics, stochastic perturbations with finite coherence time, and the nonlinear evolution after cloud saturation. For dense, thick, magnetized, or self-gravitating disks, a more self-consistent treatment of disk gravity and backreaction will also be required. The central physical conclusion remains that realistic accretion environments can redistribute gravitational-atom populations among growing and decaying modes, thereby modifying the effective superradiant evolution and the interpretation of ultralight-boson searches around astrophysical BHs.

\begin{acknowledgments}
We thank Chikako Idegawa for carefully reading the manuscript and for helpful checks and corrections. This work is supported by the National Natural Science Foundation of China (NNSFC) Grant No.12475111, No. 12205387, the Undergraduate Research Innovation Project, and the Fundamental Research Funds for the Central Universities, Sun Yat-sen University.
\end{acknowledgments}

\appendix

\renewcommand{\theequation}{\Alph{section}\arabic{equation}}

\section{Effective growth rate}
\label{A}
\setcounter{equation}{0}

According to Eq.~\eqref{growth_rate}, each eigenstate $\ket{n\ell m}$ is associated with a level-dependent growth/decay rate $\Gamma_{n\ell m}$. It is therefore convenient to introduce a rate operator $\hat\Gamma$ defined by
\begin{equation}
    \hat\Gamma\ket{n\ell m}=\Gamma_{n\ell m}\ket{n\ell m}.
\end{equation}
The states $\ket{n\ell m}$ are eigenstates of $\hat\Gamma$. For a general superposition,
\begin{equation}
    \ket{\psi}=\sum_{n,\ell,m}c_{n\ell m}\ket{n\ell m},
\end{equation}
an effective growth rate is defined as the expectation value of $\hat\Gamma$,
\begin{equation}
    \Gamma_{\mathrm{eff}}
    =\braket{\psi|\hat\Gamma|\psi}
    =\sum_{n,\ell,m}|c_{n\ell m}|^2\Gamma_{n\ell m}.
\end{equation}

\section{Timescale for the evolution of the horizon angular velocity}
\label{B}
\setcounter{equation}{0}

Accretion proceeds through the inner disk and ultimately feeds the BH from the vicinity of ISCO~\cite{bardeen1970kerr}. Both the BH mass $M$ and angular momentum $J$ therefore evolve, which in turn changes the horizon angular velocity $\Omega_H$~\cite{thorne1974disk,Shapiro:1983du}.

The horizon angular velocity is given by
\begin{equation}
    \Omega_H=\frac{a}{2M r_+}
    =\frac{\tilde a}{2M(1+X)},
    \qquad
    X=\sqrt{1-\tilde a^2},
\end{equation}
where $\tilde a\equiv a/M$ is the dimensionless spin parameter. Taking a logarithmic time derivative yields
\begin{equation}
    \frac{\dot\Omega_H}{\Omega_H}
    =\diff{}{t}\ln\Omega_H
    =\frac{\dot{\tilde a}}{\tilde a}-\frac{\dot M}{M}-\frac{1}{1+X}\dot X .
\end{equation}
Using $X=\sqrt{1-\tilde a^2}$ together with $\tilde a=J/M^2$, this expression can be rewritten as
\begin{equation}
    \frac{\dot\Omega_H}{\Omega_H}
    =\frac{1}{X}\left(\frac{\dot J}{J}-\frac{2\dot M}{M}\right)-\frac{\dot M}{M}.
\end{equation}
The angular momentum inflow is related to the mass inflow through the specific angular momentum at the ISCO,
\begin{equation}
    \dot J = l_{\mathrm{isco}}\,\dot M,
\end{equation}
so that
\begin{equation}
    \frac{\dot J}{J}
    =\frac{l_{\mathrm{isco}}}{\tilde a M}\frac{\dot M}{M}.
\end{equation}
Defining $\tilde l_{\mathrm{isco}}\equiv l_{\mathrm{isco}}/M$, the characteristic time scale for the evolution driven by accretion of $\Omega_H$ becomes
\begin{equation}
    \tau_{\Omega}
    \equiv
    \frac{\Omega_H}{|\dot\Omega_H|}
    =
    \left|
    \frac{\tilde a\sqrt{1-\tilde a^2}}
    {\tilde l_{\mathrm{isco}}-\tilde a\left(2-\sqrt{1-\tilde a^2}\right)}
    \right|
    \frac{M}{\dot M}.
\end{equation}

In the above derivation, the ISCO specific angular momentum $l_{\mathrm{isco}}$ is obtained from~\cite{Chandrasekhar:1984siy}
\begin{equation}
    l=\frac{M^{1/2}\left(r^2\mp 2aM^{1/2}r^{1/2}+a^2\right)}
    {r^{3/4}\left(r^{3/2}-3Mr^{1/2}\pm 2aM^{1/2}\right)^{1/2}},
\end{equation}
evaluated at
\begin{align}
    r_{isco}&=M\left(3+Z_2\mp\sqrt{(3-Z_1)(3+Z_1+2Z_2)}\right),\\
    \notag Z_1&= 1 + (1 - \tilde a^2)^{1/3}\left[(1 + \tilde a)^{1/3} + (1 -\tilde a)^{1/3}\right],\\
    \notag Z_2&= \sqrt{3\tilde a^2 + Z_1^2}.
\end{align}

\section{Spherical harmonics and symmetry constraints}
\label{C}
\setcounter{equation}{0}

The spherical harmonics are defined as
\begin{equation}
    Y_{\ell m}(\theta,\phi)
    =N_{\ell m}P^{m}_\ell(\cos\theta)\,e^{\i m\phi}
    =\sqrt{\frac{2\ell+1}{4\pi}\frac{(\ell-m)!}{(\ell+m)!}}\,
    P^{m}_\ell(\cos\theta)\,e^{\i m\phi},
    \qquad m\in\mathbb N,
\end{equation}
where $N_{\ell m}$ is a normalization constant and $P^{|m|}_\ell$ denotes the associated Legendre polynomials. Useful identities include
\begin{equation}
    Y_{\ell m}(\pi-\theta,\phi)=(-1)^{\ell+m}Y_{\ell m}(\theta,\phi),
    \label{character1}
\end{equation}
and
\begin{equation}
    Y_{\ell,-m}(\theta,\phi)=Y_{\ell m}^\ast(\theta,\phi)=Y_{\ell m}(\theta,-\phi).
    \label{character2}
\end{equation}

Consider the integral
\begin{align}
    \notag I_{\ell m}^{m^\prime}
    &=\int_{0}^{2\pi}Y_{\ell m}(\pi/2,\phi)\,e^{-\i m^\prime\phi}\,\mathrm{d}\phi\\
    &=N_{\ell m}P^{m}_\ell(0)\int_{0}^{2\pi}e^{\i(m-m^\prime)\phi}\,\mathrm{d}\phi .
\end{align}
The associated Legendre polynomial at zero satisfies
\begin{equation}
     P_{\ell}^m(0)=
    \begin{cases}
        0 & \text{$\ell+m$ is odd},\\[4pt]
        \displaystyle (-1)^{(\ell+m)/2}\frac{(\ell+m-1)!!}{(\ell-m)!!} & \text{$\ell+m$ is even},
    \end{cases}
\end{equation}
and
\begin{equation}
     \int_{0}^{2\pi}e^{\i(m-m^\prime)\phi}\,\mathrm{d}\phi=2\pi\,\delta_{mm^\prime}.
\end{equation}
Therefore, $I_{\ell m}^{m^\prime}$ is nonzero only if $m=m^\prime$ and $\ell+m$ is even.

For an equatorial disk, the source multipoles entering Eq.~\eqref{Id} take the form
\begin{equation}
    I_d^{(m_d)}
    =\sum_m\int_{r_{\mathrm{in}}}^{r_{\mathrm{out}}}\frac{\Sigma_m(t,r^\prime)}{{r^\prime}^2}
    \int_{0}^{2\pi}Y^\ast_{2m_d}(\pi/2,\phi)\,e^{\i m\phi}\,\mathrm{d}\phi
    \equiv\sum_{m}\tilde\Sigma_m\cdot (I_{2 m_d}^{m})^\ast,
    \label{integration}
\end{equation}
where $\tilde\Sigma_m$ denotes the corresponding radial integral. For Eq.~\eqref{integration} to be nonzero, one requires $2+m_d$ to be even, i.e.\ $m_d$ even. Combined with the selection rule, this implies that $m-m^\prime$ must be even, so $\ket{210}$ does not couple to the other two sublevels within the $\ell_d=2$ equatorially symmetric quadrupole perturbation.

This decoupling can also be understood from equatorial-reflection parity. Define the equatorial reflection operator $\mathcal P_{\mathrm{eq}}$ by
\begin{equation}
     \mathcal P_{\mathrm{eq}}:\ \theta\to\pi-\theta,\qquad \phi\to\phi.
\end{equation}
If the disk is symmetric under equatorial reflection, then
\begin{equation}
     \mathcal P_{\mathrm{eq}}V_2\mathcal P_{\mathrm{eq}}^{-1}=V_2,
\end{equation}
so $V_2$ is even under $\mathcal P_{\mathrm{eq}}$. Using Eq.~\eqref{character1}, one finds
\begin{align*}
     \mathcal P_{\mathrm{eq}}\ket{211}&=\ket{211},\\
     \mathcal P_{\mathrm{eq}}\ket{210}&=-\ket{210},\\
     \mathcal P_{\mathrm{eq}}\ket{21-1}&=\ket{21-1}.
\end{align*}
Thus $\ket{211}$ and $\ket{21-1}$ are even-parity states, while $\ket{210}$ is odd. As a representative example,
\begin{equation}
     \bra{211}V_2\ket{210}
     =\bra{211}\mathcal P_{\mathrm{eq}}^{-1}\mathcal P_{\mathrm{eq}}V_2\mathcal P_{\mathrm{eq}}^{-1}\mathcal P_{\mathrm{eq}}\ket{210}
     =-\bra{211}V_2\ket{210},
\end{equation}
so the matrix element vanishes identically.

{\color{blue}\section{Parameter estimates for $\eta_P$ and $\eta_Q$ in the two models}
\label{app:eta}

\subsection{Gaussian-envelope spiral density-wave model}
\label{subapp:spiral}

In this model, the original $P$ subspace is reduced to a two-level subspace,
\begin{equation}
    P^\prime=\mathrm{span}\{\ket{211},\ket{21-1}\}.
\end{equation}
We denote $\ket{g}\equiv\ket{211}$ and $\ket{d}\equiv\ket{21-1}$.
The effective energy separation between the two levels is
\begin{equation}
    \Delta E_{gd}=2\epsilon_h-\Omega_d .
\end{equation}
Therefore, the mixing parameter for this model can be defined as
\begin{equation}
    \eta_P^{\rm spiral}\equiv
    \left|
    \frac{9\kappa\tilde{\Sigma}_2(t)}
    {2\epsilon_h-\Omega_d}
    \right| .
\end{equation}
This definition is directly related to the control parameter $\Lambda$ used in
the main text. In particular, near the peak of the wave packet,
\begin{equation}
    \eta_{P,\rm peak}^{\rm spiral}
    =
    \left|
    \frac{9\kappa\tilde{\Sigma}_2(t\simeq0)}
    {2\epsilon_h-\Omega_d}
    \right|
    =
    \frac{\Omega_{\rm mix}(t\simeq0)}
    {|2\epsilon_h-\Omega_d|}
    \equiv
    \Lambda .
\end{equation}
Substituting the Gaussian-packet estimate for $\tilde{\Sigma}_2(t)$ in
Eq.~\eqref{eq:tildeSigma2}, we obtain
\begin{equation}
    \eta_P^{\rm spiral}(t)
    \simeq
    \frac{18\pi M}{\alpha^3|2\epsilon_h-\Omega_d|}
    \sqrt{\frac{\pi}{2}}\,
    \frac{r_0\sigma_r A}{R(t)^3}
    e^{-k_r^2\sigma_r^2/2}.
\end{equation}
Introducing the dimensionless variables
\begin{equation}
    \bar r_0=\frac{r_0}{M},
    \quad
    \bar\sigma_r=\frac{\sigma_r}{M},
    \quad
    \bar R(t)=\frac{R(t)}{M},
    \quad
    \bar k_r=k_rM,
    \quad
    \widehat{\Omega}_d=M\Omega_d ,
    \quad
    \bar\epsilon_h=M\epsilon_h=\frac{1}{12}\tilde a\alpha^6 ,
\end{equation}
the above expression becomes
\begin{equation}
    \eta_P^{\rm spiral}(t)
    \simeq
    \frac{18\pi(AM)}
    {\alpha^3\left|\tilde a\alpha^6/6-\widehat{\Omega}_d\right|}
    \sqrt{\frac{\pi}{2}}\,
    \frac{\bar r_0\bar\sigma_r}{\bar R(t)^3}
    e^{-\bar k_r^2\bar\sigma_r^2/2}.
\end{equation}
Taking $t\simeq0$ gives the maximum value. Substituting the parameters used in
this work,
\begin{equation}
    \bar r_0=4,
    \qquad
    \bar\sigma_r=1,
    \qquad
    \bar k_r=1,
    \qquad
    \tilde a=0.9,
    \qquad
    \widehat{\Omega}_d=10^{-9},
\end{equation}
we find
\begin{equation}
    \eta_{P,\rm peak}^{\rm spiral}
    \simeq
    17.91\times(AM)
    \frac{1}
    {\alpha^3\left|\alpha^6-6.67\times 10^{-9}\right|}
    \equiv
    17.91(AM)\times F(\alpha).
\end{equation}
In the scan range $\alpha\in[0.03,0.05]$ considered here, if one takes $A\simeq\Sigma_0$, the mixing parameter satisfies $\eta_P^{\rm spiral}\ll1$ everywhere except in the vicinity of the resonance point $\alpha_{\rm res}\simeq0.0434$. To ensure compatibility across all scenarios, we directly solve the Schr\"odinger equation in the interaction picture.

We now estimate the leakage parameter $\eta_Q$ into external states. For an
arbitrary external state $\ket{a}=\ket{n'\ell'm'}\in Q$ and a state
$\ket{21m}$ inside the three-level subspace, the quadrupolar perturbation
matrix element can be written as
\begin{equation}
    \bra{a}V_{\rm spiral}(t)\ket{21m}
    =
    -\frac{4\pi\mu}{5}
    \sum_{m_d}
    I_d^{(m_d)}(t)
    \left[
    \int_0^\infty \d r\,r^4 R_{n'\ell'}R_{21}
    \right]
    \left[
    \int \d\Omega\,
    Y_{\ell'm'}^\ast Y_{2m_d}Y_{1m}
    \right].
\end{equation}
The first bracket is the radial overlap integral, while the second bracket is
the angular integral. The characteristic scale of the radial overlap integral
is controlled by the Bohr radius $a=M/\alpha^2$. For low-lying external states,
one has
\begin{equation}
    \int_0^\infty \d r\,r^4 R_{n'\ell'}(r)R_{21}(r)
    \sim
    a^2\times \mathcal O(10).
\end{equation}
The angular integral only imposes selection rules and contributes an
$\mathcal O(1)$ dimensionless coefficient. Therefore, the external-state matrix
element has the same parametric order as the quadrupolar matrix element within
the three-level subspace, up to a dimensionless coefficient
$C_Q^{\rm spiral}=\mathcal O(1\sim10)$, namely
\begin{equation}
    |\bra{a}V_{\rm spiral}(t)\ket{21m}|
    \sim
    C_Q^{\rm spiral}\kappa|\tilde{\Sigma}_2(t)|.
\end{equation}
For a conservative estimate, we take $C_Q^{\rm spiral}\simeq10$, corresponding
to the assumption that the coupling to external states is not smaller than the
coupling scale of the internal $\ket{211}\leftrightarrow\ket{21-1}$ channel.

The leakage parameter into external states can then be estimated as
\begin{equation}
    \eta_Q^{\rm spiral}(t)
    \equiv
    \max_{a\in Q,\ i\in P}
    \left|
    \frac{\bra{a}V_{\rm sp}(t)\ket{i}}
    {E_a-E_i-\Omega_d}
    \right|
    \sim
    \frac{C_Q\kappa|\tilde{\Sigma}_2(t)|}
    {\Delta_Q},
\end{equation}
where
\begin{equation}
    \Delta_Q
    \equiv
    \min_{a\in Q,\ i\in P}
    |E_a-E_i-\Omega_d|
\end{equation}
is the minimum effective energy separation between the $P$ subspace and the
allowed external states. For the low-frequency spiral wave considered in this
work, $\Omega_d$ is much smaller than the spacing to external principal levels,
so we may approximate
\begin{equation}
    \Delta_Q
    \simeq
    \min_{a\in Q,\ i\in P}|E_a-E_i|.
\end{equation}
Since the leading quadrupolar perturbation does not directly couple
$\ket{21m}$ to $\ket{200}$ at first order, a conservative estimate of the
external energy gap can be obtained from the nearest allowed external bound
state with $n=3,\ell=1$, giving
\begin{equation}
    \Delta_Q
    \simeq
    |E_3-E_2|
    =
    \frac{5}{72}\frac{\alpha^3}{M}.
\end{equation}
Thus,
\begin{equation}
    \eta_Q^{\rm spiral}(t)
    \sim
    \frac{144\pi C_Q^{\rm spiral}}{5}
    (AM)\alpha^{-6}
    \sqrt{\frac{\pi}{2}}\,
    \frac{\bar r_0\bar\sigma_r}{\bar R(t)^3}
    e^{-\bar k_r^2\bar\sigma_r^2/2}.
\end{equation}
Taking $t\simeq0$ gives the maximum value. Substituting the parameters used in
this work, we find
\begin{equation}
    \eta_{Q,\rm peak}^{\rm spiral}
    \sim
    1.2\times10^3\,(AM)\alpha^{-6}.
\end{equation}
If $A\simeq\Sigma_0$ is adopted, then within the parameter range scanned in this
work, $\eta_Q^{\rm spiral}$ remains much smaller than unity. Therefore, although
the spiral wave can induce $\mathcal O(1)$ or even stronger mixing within the
three-level subspace, leakage into the $Q$ subspace remains perturbatively small
because the external-state energy gap $\Delta_Q\sim\mu\alpha^2$ is much larger
than the internal three-level splitting $\epsilon_h\sim\mu\tilde a\alpha^5$.

\subsection{Warped accretion-disk model}
\label{subapp:warp}

In the quasi-static warped accretion-disk model, the three-level subspace is
\begin{equation}
    P=\mathrm{span}\{\ket{211},\ket{210},\ket{21-1}\}.
\end{equation}
The perturbation opens two internal mixing channels within this three-level
subspace,
$\ket{211}\leftrightarrow\ket{210}$ and
$\ket{210}\leftrightarrow\ket{21-1}$, whose effective energy separations are
respectively
\begin{equation}
    \Delta_{g0}=\epsilon_h-9\kappa\tilde{\Sigma}_0,
    \qquad
    \Delta_{0d}=\epsilon_h+9\kappa\tilde{\Sigma}_0 .
\end{equation}
Therefore, the internal three-level mixing parameter in the warped-disk model
can be defined as
\begin{equation}
    \eta_P^{\rm warp}
    \equiv
    \max\left[
    \frac{\overline{\Sigma}}{|\Delta_{g0}|},
    \frac{\overline{\Sigma}}{|\Delta_{0d}|}
    \right],
\end{equation}
namely
\begin{equation}
    \eta_P^{\rm warp}
    =
    \max\left[
    \frac{\frac{9\sqrt{2}}{2}\kappa|\tilde{\Sigma}|}
    {|\epsilon_h-9\kappa\tilde{\Sigma}_0|},
    \frac{\frac{9\sqrt{2}}{2}\kappa|\tilde{\Sigma}|}
    {|\epsilon_h+9\kappa\tilde{\Sigma}_0|}
    \right].
\end{equation}
This definition shows that the internal three-level mixing induced by the
warped disk depends not only on the off-diagonal coupling
$\tilde{\Sigma}$, but also on whether the diagonal disk potential drives a
pair of levels close to degeneracy.

Using the warped-disk model and parameter values adopted in this work, we find
$R_{\rm BP}\simeq2.01\times10^3 M$. Taking
$x_{\rm in}=2.32$ and $x_{\rm out}=10^5$, we obtain
\begin{equation}
    M^2\tilde{\Sigma}_0
    \simeq
    1.31\times10^{-16},
    \qquad
     M^2\tilde{\Sigma}
    \simeq
    \left(2.42-0.69i\right)\times10^{-17}.
\end{equation}
The corresponding weighted effective tilt angle is therefore
\begin{equation}
    \beta_{\rm eff}
    \equiv
    \frac{|\tilde{\Sigma}|}{\tilde{\Sigma}_0}
    \simeq
    0.192 .
\end{equation}

To display the parameter dependence of $\eta_P^{\rm warp}$ more transparently,
we define
\begin{equation}
    X(\alpha)
    \equiv
    \frac{9\kappa\tilde{\Sigma}_0}{\epsilon_h}
    =
    \frac{216\pi}{\tilde a}
    M^2\tilde{\Sigma}_0
    \alpha^{-9}
    \simeq
    9.88\times10^{-14}\alpha^{-9}.
\end{equation}
Then
\begin{equation}
    \eta_P^{\rm warp}
    =
    \max\left[
    \frac{\beta_{\rm eff}}{\sqrt{2}}
    \frac{X(\alpha)}{|1-X(\alpha)|},
    \frac{\beta_{\rm eff}}{\sqrt{2}}
    \frac{X(\alpha)}{|1+X(\alpha)|}
    \right].
\end{equation}
When $X(\alpha)\simeq1$, one has
$\Delta_{g0}=\epsilon_h-9\kappa\tilde{\Sigma}_0\simeq0$. In this case, the
effective energy separation between $\ket{211}$ and $\ket{210}$ is suppressed,
and $\eta_P^{\rm warp}$ is significantly enhanced. Solving $X(\alpha)=1$ gives
the near-degeneracy point
\begin{equation}
    \alpha_{\rm res}
    \simeq
    0.0359 .
\end{equation}
Therefore, near $\alpha_{\rm res}$, the three-level internal mixing induced by
the warped disk can become non-perturbatively strong; away from this
near-degeneracy point, the mixing strength rapidly decreases.

We now estimate the external-state leakage parameter $\eta_Q$. Since the
warped-disk perturbation is quasi-static, the leakage parameter is defined as
\begin{equation}
    \eta_Q^{\rm warp}
    \equiv
    \max_{a\in Q,\ i\in P}
    \left|
    \frac{\bra{a}V_{\rm warp}\ket{i}}
    {E_a-E_i}
    \right| .
\end{equation}
For an arbitrary external state $\ket{a}=\ket{n'\ell'm'}\in Q$, following the
same reasoning as in the previous subsection, we obtain
\begin{equation}
    |\bra{a}V_{\rm warp}\ket{21m}|
    \sim
    C_Q^{\rm warp}\kappa|\tilde{\Sigma}|,
    \qquad
    C_Q^{\rm warp}=\mathcal{O}(1\sim10).
\end{equation}
To be consistent with the conservative estimate used for the spiral-wave model,
we take $C_Q^{\rm warp}\simeq10$. The leakage parameter can then be estimated as
\begin{equation}
    \eta_Q^{\rm warp}
    \sim
    \frac{C_Q^{\rm warp}\kappa|\tilde{\Sigma}|}{\Delta_Q}
    =
    \frac{144\pi C_Q^{\rm warp}}{5}
    M^2|\tilde{\Sigma}|\alpha^{-6}
    \simeq2.05\times10^{-14}\alpha^{-6}.
\end{equation}

In the parameter range considered in this work,
$\eta_Q^{\rm warp}\sim\mathcal O(10^{-6}\sim10^{-5})$, namely
$\eta_Q^{\rm warp}\ll1$. Therefore, leakage from the warped-disk perturbation
into the $Q$ subspace can be neglected, and the three-level truncation is
self-consistent. On the other hand, $\eta_P^{\rm warp}$ is significantly
enhanced near $\alpha\simeq\alpha_{\rm res}$. Around
$\alpha_{\rm res}\simeq0.0359$, the diagonal splitting $\Delta_{g0}$ is
suppressed by the axisymmetric disk potential, so $\eta_P^{\rm warp}$ can reach
$\mathcal O(1)$ or larger. In this regime, the non-degenerate perturbative
expansion inside the three-level subspace is no longer reliable. One should
instead diagonalize the near-degenerate $\ket{211}$--$\ket{210}$ subblock, or
directly solve the full three-level effective Hamiltonian. If
$\eta_P^{\rm warp}$ is used as the criterion for the clear breakdown of
non-degenerate perturbation theory, the approximate breakdown interval is
$0.0353\lesssim\alpha\lesssim0.0364$.}

\section{Damping timescale for warp precession}
\label{D}
\setcounter{equation}{0}

Consider a geometrically thin Keplerian disk with a small tilt. The
Lense--Thirring precession frequency is~\cite{Bardeen:1975zz}
\begin{equation}
     \Omega_{\mathrm{LT}}(r)=\frac{2\tilde a M^2}{r^3}.
\end{equation}
Let $\bm L(r,t)$ denote the angular-momentum surface density~\cite{1996MNRAS.282..291S}. In the
Papaloizou--Pringle approximation, if internal communication in the disk is sufficiently efficient, the warped disk can be approximately treated as undergoing rigid-body precession~\cite{10.1093/mnras/202.4.1181}.  We take the magnitude profile of the angular-momentum surface density to be steady, while allowing only its direction to evolve:
\begin{equation}
    \bm L(r,t)=L(r)\bm l(t),
    \qquad
    L(r)=\Sigma(r)r^2\Omega_K(r),
    \qquad
    \Omega_K(r)=\sqrt{\frac{M}{r^3}}.
\end{equation}
\color{blue}
Then, in the rigid-precession approximation, the global mean precession frequency is defined as
\begin{equation}
     \Omega_p
     =
     \frac{
     \int_{r_{\mathrm{in}}}^{r_{\mathrm{out}}}
     \Omega_{\mathrm{LT}}(r)\Sigma(r)r^3\Omega_K(r)\,\mathrm d r
     }{
     \int_{r_{\mathrm{in}}}^{r_{\mathrm{out}}}
     \Sigma(r)r^3\Omega_K(r)\,\mathrm d r
     }.
\end{equation}

To obtain an explicit estimate, we assume a power-law surface-density profile,
\begin{equation}
    \Sigma(r)=\Sigma_0\left(\frac{r}{M}\right)^{-p}.
\end{equation}
This gives
\begin{equation}
     \Omega_p
     =
     2\tilde a M^2
     \frac{
     \int_{r_{\mathrm{in}}}^{r_{\mathrm{out}}}r^{-p-3/2}\,\mathrm d r
     }{
     \int_{r_{\mathrm{in}}}^{r_{\mathrm{out}}}r^{3/2-p}\,\mathrm d r
     }.
\end{equation}
Defining $x\equiv r_{\mathrm{out}}/r_{\mathrm{in}}$, for $p\neq -1/2$ and $p\neq 5/2$ one obtains
\begin{equation}
     \Omega_p
     =
     \Omega_{\mathrm{LT}}(r_{\mathrm{in}})F_p(x),
\end{equation}
where
\begin{equation}
     F_p(x)
     =
     \frac{5/2-p}{p+1/2}
     \frac{
     1-x^{-(p+1/2)}
     }{
     x^{5/2-p}-1
     }.
\end{equation}
\color{black}

We next estimate the damping timescale of the precession. We identify the damping time, at the order-of-magnitude level, with the warp-diffusion time across a characteristic radius $R_\ast$:
\begin{equation}
     \tau_{\mathrm{damp}}\sim\frac{R_\ast^2}{\nu_2}.
\end{equation}
Using $\nu_2=\alpha_2 c_s H,\  c_s=\Omega_K H$, and defining $h\equiv H/r$, we obtain~\cite{ogilvie1999non}
\begin{equation}
     \tau_{\mathrm{damp}}
     \sim
     \frac{1}{\alpha_2h^2}\frac1{\Omega_K(R_\ast)}.
\end{equation}
In the Papaloizou--Pringle approximation, one may take $\alpha_2\sim 1/(2\alpha)$.

\color{blue}
Taking the ratio between the damping timescale and the precession period gives, up to factors of order unity,
\begin{equation}
     \frac{\tau_{\mathrm{damp}}}{T_p}
     \sim
     \frac{1}{\alpha_2}
     \frac{\Omega_p}{\Omega_K(R_\ast)}
     \left(
     \frac{R_\ast}{H(R_\ast)}
     \right)^2
     =
     \frac{2\tilde a}{\alpha_2 h^2}
     F_p(x)
     \left(\frac{M}{r_{\mathrm{in}}}\right)^{3/2}
     \left(\frac{R_\ast}{r_{\mathrm{in}}}\right)^{3/2}.
\end{equation}
If we further take the characteristic radius to be
$R_\ast\sim \sqrt{r_{\mathrm{in}}r_{\mathrm{out}}}$, then
\begin{equation}
     \frac{\tau_{\mathrm{damp}}}{T_p}
     \sim
     \frac{2\tilde a}{\alpha_2 h^2}
     F_p(x)
     \left(\frac{M}{r_{\mathrm{in}}}\right)^{3/2}
     x^{3/4}.
     \label{precession}
\end{equation}
This is the self-consistent order-of-magnitude estimate adopted in this work.
\color{black}

\section{Geometric model for a warped disk}
\label{E}
\setcounter{equation}{0}

A geometrically thin warped disk can be modeled as a collection of rings. A ring at radius $r$ is obtained from the corresponding equatorial ring by a rotation about the $y$ axis by $\beta(r)$ followed by a rotation about the $z$ axis by $\gamma(r)$, with $\beta\ll1$.

Consider a point on the equatorial ring,
\begin{equation}
    P_0=(r\cos\phi,r\sin\phi,0)^T.
\end{equation}
After the two rotations, the corresponding point on the warped ring is
\begin{equation}
     P=R_z(\gamma)R_y(\beta)P_0=
     \begin{pmatrix}
        r\cos\phi\cos\gamma-r\sin\phi\cos\beta\sin\gamma\\
        r\cos\phi\sin\gamma+r\sin\phi\cos\beta\cos\gamma\\
        r\sin\phi\sin\beta
     \end{pmatrix},
     \label{P}
\end{equation}
where
\begin{equation}
     R_y(\beta)=
     \begin{pmatrix}
         \cos\beta & 0 & -\sin\beta \\
         0 & 1 & 0 \\
         \sin\beta & 0 & \cos\beta
     \end{pmatrix},
     \qquad
     R_z(\gamma)=
     \begin{pmatrix}
         \cos\gamma & -\sin\gamma & 0 \\
         \sin\gamma & \cos\gamma & 0 \\
         0 & 0 & 1 
     \end{pmatrix}.
\end{equation}
Expanding Eq.~\eqref{P} to leading order in $\beta$ and rewriting it in spherical coordinates gives
\begin{equation}
     P\simeq
     \begin{pmatrix}
        r\cos(\phi+\gamma)\\
        r\sin(\phi+\gamma)\\
        r\beta\cos\phi
     \end{pmatrix}
     =(r,\phi+\gamma,\arccos(\beta\cos\phi)).
\end{equation}
Therefore, when evaluating the warped-disk contribution to the gravitational potential, the equatorial expression can be used with the substitutions
\begin{equation}
    \phi\to\phi+\gamma,
    \qquad
    \theta\to\arccos(\beta\cos\phi).
\end{equation}

Assuming an axisymmetric surface density $\Sigma=\Sigma(t,r^\prime)$, the azimuthal integral appearing in Eq.~\eqref{Id} becomes
\begin{equation}
     I_\phi^{(m_d)}=\int_{0}^{2\pi}Y_{2m_d}^\ast(\arccos(\beta\cos\phi^\prime),\phi^\prime+\gamma)\,\mathrm{d}\phi^\prime.
\end{equation}
To the lowest relevant order in $\beta$,
\begin{align}
    I_\phi^{(\pm2)}&=-\frac 18\sqrt{\frac{15\pi}{2}}\beta^2 \,e^{\mp2i\gamma},\\
    I_\phi^{(\pm1)}&=\mp\frac 12\sqrt{\frac{15\pi}{2}}\beta \,e^{\mp i\gamma},\\
    I_\phi^{(0)}&=\frac 14\sqrt{5\pi}(3\beta^2-2).
\end{align}
Thus $I_\phi^{(\pm2)}=\mathcal O(\beta^2)$ while $I_\phi^{(\pm1)}=\mathcal O(\beta)$. Retaining only terms linear in the small tilt implies that the $m_d=\pm2$ contributions can be neglected at this order; by the selection rule, the $\ket{211}\leftrightarrow\ket{21-1}$ matrix element vanishes to $\mathcal O(\beta)$. In addition, $I_\phi^{(0)}\simeq-\sqrt{5\pi}/2$ at leading order.

{\color{blue}\section{Periodically Precessing Warp}
\label{app:precessing_warp}

In the main text, our discussion of the disk-induced perturbations assumes a quasi-static warp. As a complementary coherent-driving limit, this appendix considers the scenario of a periodically precessing warp. We note that such a configuration should not be regarded as the generic outcome of a freely evolving large-scale thin disk; rather, its realization typically requires either a localized precessing region or continuous support from external torques, disk self-gravity, or a global disk mode. Under this periodic driving, the primary physical consequence of the precession is to shift the resonance conditions within the retained $\ket{21m}$ subspace, as detailed below.

To make this coherent-driving limit explicit, we model the time-dependent precessing twist phase as
\begin{equation}
\gamma(r,t)=\gamma_0(r)+s\Omega_p t,
\qquad
s=\pm1,
\end{equation}
where $s$ denotes the direction of precession and $\Omega_p$ is the precession frequency. Under this ansatz, the weighted warp moment defined in the main text gains an oscillatory phase factor:
\begin{equation}
\widetilde\Sigma(t)=\widetilde\Sigma_{\rm w}e^{-\i s\Omega_p t},
\end{equation}
where the time-independent amplitude $\widetilde\Sigma_{\rm w}$ is evaluated as the radial integral
\begin{equation}
\widetilde\Sigma_{\rm w}
=
\int_{r_{\rm in}}^{r_{\rm out}}
\frac{\Sigma(r')}{r'^2}
\beta(r')e^{-\i\gamma_0(r')}\mathrm{d} r' .
\end{equation}
Consequently, the static warp Hamiltonian is promoted to a time-dependent driving Hamiltonian within the $\{|211\rangle, |210\rangle, |21{-}1\rangle\}$ basis:
\begin{equation}
H_{\rm prec}(t)=
\begin{pmatrix}
E_g & -\bar\Sigma_{\rm w}e^{-\i s\Omega_p t} & 0\\
-\bar\Sigma_{\rm w}^*e^{\i s\Omega_p t} & E_0
& \bar\Sigma_{\rm w}e^{-\i s\Omega_p t}\\
0 & \bar\Sigma_{\rm w}^*e^{\i s\Omega_p t} & E_d
\end{pmatrix},
\end{equation}
where the diagonal unperturbed energies modified by the symmetric warp component $\widetilde\Sigma_0$ are given by
\begin{equation}
E_g=E+\epsilon_h-3\kappa\widetilde\Sigma_0,
\qquad
E_0=E+6\kappa\widetilde\Sigma_0,
\qquad
E_d=E-\epsilon_h-3\kappa\widetilde\Sigma_0,
\end{equation}
and the effective coupling strength is redefined as
\begin{equation}
\bar\Sigma_{\rm w}
=
\frac{9\sqrt{2}}{2}\kappa\widetilde\Sigma_{\rm w}.
\end{equation}

In this coherent limit, and neglecting state damping during a single precession-driven mixing episode, the explicit time dependence of $H_{\rm prec}(t)$ can be removed by moving into a co-rotating frame. We introduce the unitary rotating-frame transformation for the state coefficients:
\begin{equation}
\begin{pmatrix}
c_g \\ c_0 \\ c_d
\end{pmatrix}
=
\mathrm{diag}
\left(e^{-\i s\Omega_p t},1,e^{\i s\Omega_p t}\right)
\begin{pmatrix}
a_g \\ a_0 \\ a_d
\end{pmatrix}.
\end{equation}
Substituting this transformation into the projected Schr\"odinger equation yields the effective static Hamiltonian in the rotating frame:
\begin{equation}
H_{\rm rot}=
\begin{pmatrix}
E_g-s\Omega_p & -\bar\Sigma_{\rm w} & 0\\
-\bar\Sigma_{\rm w}^* & E_0 & \bar\Sigma_{\rm w}\\
0 & \bar\Sigma_{\rm w}^* & E_d+s\Omega_p
\end{pmatrix}.
\end{equation}
From the structure of $H_{\rm rot}$, it is evident that a periodic precession does not alter the $m_d=\pm1$ selection channel opened by the warp geometry; its sole effect is to shift the diagonal energies by $\pm \Omega_p$ depending on the precession direction $s$. 

Accordingly, the static near-degeneracy conditions discussed in the main text are replaced by the frequency-matching resonance conditions:
\begin{align}
\label{eq:prec_res1}
\ket{211}\leftrightarrow\ket{210}:
\qquad
\Delta_{g0} &\simeq s\Omega_p,
\\
\label{eq:prec_res2}
\ket{210}\leftrightarrow\ket{21{-}1}:
\qquad
\Delta_{0d} &\simeq s\Omega_p,
\end{align}
where the real energy splittings are defined as
\begin{align}
\Delta_{g0} &\equiv E_g-E_0=\epsilon_h-9\kappa\widetilde\Sigma_0, \\
\Delta_{0d} &\equiv E_0-E_d=\epsilon_h+9\kappa\widetilde\Sigma_0.
\end{align}
Crucially, for a fixed precession frequency $\Omega_p$ and direction $s$, the two resonance conditions in Eqs.~\eqref{eq:prec_res1} and \eqref{eq:prec_res2} cannot generally be satisfied simultaneously due to the symmetric energy shift $9\kappa\widetilde\Sigma_0$. Therefore, a precessing warp can selectively enhance one adjacent mixing channel while leaving the other effectively unactivated.}

\bibliography{references}
\end{document}